\documentclass[pdflatex,sn-mathphys,Numbered]{sn-jnl}% Math and Physical Sciences Reference Style
\usepackage{graphicx}%
\usepackage{multirow}%
\usepackage{amsmath,amssymb,amsfonts}%
\usepackage{amsthm}%
\usepackage[title]{appendix}%
\usepackage{xcolor}%
\usepackage{textcomp}%
\usepackage{manyfoot}%
\usepackage{booktabs}%
\usepackage{algorithm}%
\usepackage{algorithmicx}%
\usepackage{algpseudocode}%
\usepackage{listings}%
\usepackage{lmodern}%
\usepackage{comment}
\usepackage{makecell} % for makecell in table
\usepackage[normalem]{ulem} % for sout
\usepackage[capitalise]{cleveref} % for cref
\usepackage{array}
\usepackage{hyperref}
\usepackage{pdfpages}
\usepackage{siunitx}    % For aligning numbers on the decimal point
\usepackage{graphicx}
\usepackage{subcaption}

\newcommand\numberthis{\addtocounter{equation}{1}\tag{\theequation}}

\newcommand{\fixme}{\textcolor{black}}

\newcommand{\edit}{\textcolor{black}}

\begin{document}

% WH: 원본: DeepBioisostere: Discovering Bioisosteres with Deep Learning for a Fine Control of Multiple Molecular Properties
% WH: 수정 제안: Autonomous Bioisosteric Replacement for Multi-Property Optimization in Drug Design
\title[Article Title]{Autonomous Bioisosteric Replacement for Multi-Property Optimization in Drug Design}


\author[1]{\fnm{Hyeongwoo} \sur{Kim}}
\equalcont{These authors contributed equally to this work.}

\author[1]{\fnm{Seokhyun} \sur{Moon}}
\equalcont{These authors contributed equally to this work.}

\author[1]{\fnm{Wonho} \sur{Zhung}}

\author[1]{\fnm{Shinwoo} \sur{Kim}}

\author[2]{\fnm{Jaechang} \sur{Lim}}

\author*[1,2,3]{\fnm{Woo Youn} \sur{Kim}}\email{wooyoun@kaist.ac.kr}

\affil[1]{\orgdiv{Department of Chemistry}, \orgname{KAIST}, \orgaddress{\street{291 Daehak-ro}, \city{Yuseong-gu}, \postcode{34141}, \state{Daejeon}, \country{Republic of Korea}}}

\affil[2]{\orgname{HITS Inc.}, \orgaddress{\street{124 Teheran-ro}, \city{Gangnam-gu}, \postcode{06234}, \state{Seoul}, \country{Republic of Korea}}}

\affil[3]{\orgdiv{Graduate School of Data Science}, \orgname{KAIST}, \orgaddress{\street{291 Daehak-ro}, \city{Yuseong-gu}, \postcode{34141}, \state{Daejeon}, \country{Republic of Korea}}}


\abstract{\edit{Optimizing molecular properties while preserving biological activity is a central challenge in drug design. Bioisosteric replacement, which substitutes a molecular fragment with a chemically or biologically analogous moiety, offers a powerful strategy for fine-tuning properties without disrupting target binding. 
However, existing in silico approaches often rely on expert-defined modification sites or suffer from modulating multiple molecular properties simultaneously.
Here, we present DeepBioisostere, a deep generative model that performs end-to-end bioisosteric replacement by autonomously selecting and substituting molecular fragments to satisfy multiple target properties. The model captures complex relationships across the molecular graph, enabling the optimization of sophisticated properties such as drug-likeness and synthetic accessibility. By learning from experimental bio-assay data, DeepBioisostere proposes replacements that maintain biological activities, even generating potential bioisosteres beyond the training data. We demonstrate the effectiveness of the model in computational hit-to-lead optimization scenarios, highlighting its potential to accelerate rational molecular design without relying on expert heuristics or pre-established substitution rules.}}

\maketitle

\section{Introduction}
\label{introduction}
\edit{Optimizing molecules to fulfill multiple properties such as bioactivity, bio-compatibility, and synthetic accessibility simultaneously is a fundamental yet challenging goal in drug design, often requiring substantial time and cost to advance a compound from the laboratory to the clinic.\cite{drugproperty}
% A promising approach to addressing this multi-objective design problem is to modify the chemical structure of a molecule that already exhibits bioactivity toward a target protein.
Given the complexity of the multi-objective design problem, a widely adopted approach is to begin with a molecule that exhibits known bioactivity toward a target protein and gradually modify its structure to enhance desirable properties.
In particular, bioisosteric replacement, the substitution of one chemical moiety with another that preserves biological activity, offers a principled way to fine-tune molecular properties with minimal disruption to the molecule's overall function.\cite{Bioisostere1,Bioisostere2, Bioisostere3} This approach allows for delicate enhancement of specific molecular properties while maintaining the essential characteristics of the original compound, requiring careful consideration of both structural and biochemical analogies. To systematize this process, many chemists have attempted to formalize common patterns observed in successful molecular replacement, namely bioisosteres. In recent decades, numerous bioisosteres have been reported, facilitating more rational and effective chemical modifications in drug development.\cite{example_tetrazole1,example_tetrazole2,phenyl-bioisosteres}}

%Optimizing molecules to fulfill multiple properties is a challenging yet fundamental goal of drug design. A drug should be bioactive, biocompatible, and synthetically accessible,\cite{drugproperty}, thereby taking an enormous amount of time and cost to bring a compound from a lab to a patient.\cite{drugcost}
%One promising approach to achieving this multi-objective molecular design problem is to modify a chemical structure from an acclaimed molecule with satisfied properties. 
%Namely, chemical modifications, such as scaffold hopping\cite{ScaffoldHopping1, ScaffoldHopping2} or R-group replacement,\cite{rgroup}, aim to delicately enhance a target property while maintaining the other properties of the original structure, where a comprehensive consideration of both structural and biochemical analogies is necessary.
%To develop a more efficient way of chemical modification, many chemists attempted to standardize useful rules that are frequently adopted in molecular optimization, leading to the emergence of the concept of bioisosteric replacement.\cite{Bioisostere1,Bioisostere2, Bioisostere3}
%Chemical moiety pairs involved in bioisosteric replacements, so-called bioisosteres, have been reported variously over the last decades and facilitated more proper chemical modifications.\cite{example_tetrazole1,example_tetrazole2,phenyl-bioisosteres}

% 기존 방법은 변환되는 쌍만을 가지고 추천을 해준다. -> 이게 문제는 아님. 뒤에서 이게 되면 더 좋다~ 라고 표현.
% 2. Similarity-based approach 소개
Traditional bioisosteric replacements have heavily relied on the intuition and experience of medicinal chemists.\cite{patani1996bioisosterism, ertl2007silico, kumari2020amide,seddon2018bioisosteric}
The identification of proper bioisosteric replacements has required extensive trial and error, leading to the rise of in silico approaches.\cite{InSilicoApproach, SwissBioisostere1}
There have been two mainstream in silico approaches to finding bioisosteres: similarity-based and database-mining methods.
Similarity-based approaches operate on the assumption that structural analogs would exhibit similar biological or chemical properties.\cite{SimilarityPropertyPrinciple}
They prioritize potential bioisosteres by comparing their structural, steric, or electronic features to those of the original moiety of interest.\cite{SimilarityApproach1, SimilarityApproach2, SimilarityApproach3, SimilarityApproach4}
Diverse molecular descriptors such as fingerprints\cite{example_similarity, example_similarity2} or structural alignments\cite{example_structure} were often employed to assess the similarity between chemical moieties.
The similarity-based methods facilitated changes in biochemical properties during a particular chemical modification prior to experimental validation.

\begin{figure*}[ht!]
 \centering
 \includegraphics[width=0.95\textwidth]{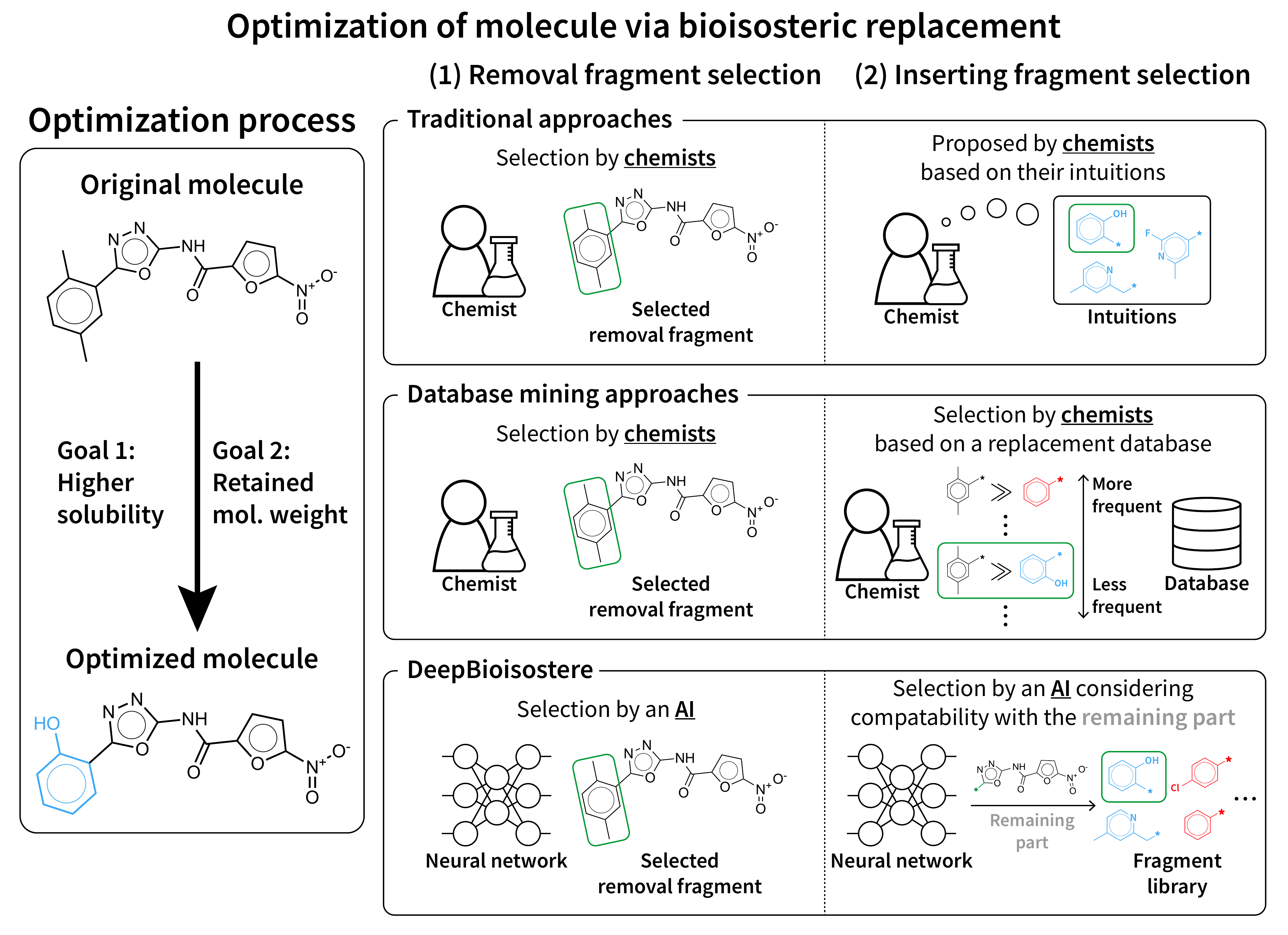}
 \caption{\textbf{An illustration of different approaches in bioisosteric replacement for given property optimization goals.} Long-standing traditional approaches have relied on the knowledge of chemists for the entire optimization process. With an appropriate removal fragment selected by a chemist, one can employ traditional database-mining approaches to think of its various potential bioisosteres. Still, the intuition of experts is required to select the proper removal moiety for the property optimization goals. Unlike the previous approaches, which inevitably depend on the intuition of experts, DeepBioisostere utilizes a neural network to mimic the sequential decision-making process of the molecule optimization process based on bioisosterism. By taking into account the entire molecular structure, DeepBioisostere can flexibly suggest modified molecules that satisfy the delicate property control conditions.}
 \label{figure:concept-figure}
\end{figure*}

On the other hand, database-mining approaches identify potential bioisosteres by extraction from large chemical databases.\cite{DatabaseMiningApproach1, DatabaseMiningApproach2, DatabaseMiningApproach3} (See \cref{figure:concept-figure})
They systematically enumerate pairs of molecules with only a minor structural difference, the so-called matched molecular pairs (MMPs),\cite{MMP} from a large bio-assay database provided in public libraries such as ChEMBL.\cite{ChEMBL}
Then, molecular fragments corresponding to the difference in each MMP are paired and identified as potential bioisosteric replacements along with their occurrences in the parent library.
The differences in chemical properties and bioactivity in biochemical assays are used to predict the effect of each bioisosteric replacement, referred to as matched molecular pairs analysis (MMPA), which is essential for subsequent bioisostere recommendations.
This statistical evidence enables the identification of a new bioisosteric replacement, even if the involved moieties are structurally unrelated at first glance.

Despite their advantages, existing approaches face limitations that hinder the identification of suitable chemical modifications. Similarity-based methods rely heavily on predefined molecular descriptors correlated with specific properties, making them dependent on expert intuition and prone to arbitrariness.\cite{NumDescriptors} They also struggle to identify bioisosteres that are structurally dissimilar but functionally equivalent. 
\edit{Meanwhile, database-mining methods are constrained by the statistical absence of uncommon or unexplored moiety pairs within the database and thus can barely identify them.
Because such methods evaluate property changes based only on the exchanged fragments, they struggle to capture complex attributes such as synthesizability or drug-likeness, which inherently depend on the entire molecular context rather than local substitutions.}
Moreover, both approaches require the manual selection of the molecular fragment to be modified, which limits fully automated molecular optimization through bioisosteric replacement.

Recently, deep generative modeling has emerged as a powerful approach for chemical modification, capturing data-intrinsic patterns that reflect both chemical structure and biological context without relying on handcrafted descriptors.\cite{FindBioWithDNN, shan2020molopt, jin2020multi, fu2021mimosa, loeffler2024reinvent, TransformerMMP1, TransformerMMP2, TransformerADMET, jin2019hierarchical, jin2020hierarchical, Modof, DeepHop, ScaffoldGVAE, maziarka2020mol}
\edit{For example, \citet{FindBioWithDNN} and \citet{shan2020molopt} applied neural networks to fragment-level data derived from ChEMBL, demonstrating the ability to identify potential bioisosteric replacements beyond the training set.}
\edit{\citet{jin2020multi}, \citet{fu2021mimosa}, and \citet{loeffler2024reinvent} employed standard optimization algorithms to refine molecular structures toward specific biological targets with low-cost labels, such as docking scores.}
\edit{\citet{DeepHop} and \citet{ScaffoldGVAE} demonstrated that deep learning methods can be effectively applied to scaffold hopping, generating alternative scaffolds while maintaining biological activity.}
While the methods mark significant progress, no existing deep learning framework for chemical modification can simultaneously control multiple properties while maintaining the original bioactivity in a general manner.
Addressing this gap is essential for fully leveraging bioisosteric replacement in automated, property-driven molecular design.

\edit{Here, we propose DeepBioisostere, a deep generative model for autonomous bioisosteric replacement with fine control over multiple molecular properties in an end-to-end manner, selecting modification sites and generating context-aware substitution. (See \cref{figure:concept-figure})
We systematically curated a standardized dataset for training with experimental bioassay data to implicitly incorporate the statistical patterns of structure-activity relationships, guiding our model’s proposed replacements toward maintaining the target activity.
%We demonstrate the compatibility of replacement with the whole molecular structure by showing the 
% Unlike similarity-based and database-mining approaches that require expert input to predefine the site of modification, DeepBioisostere learns to design the entire replacement process.
DeepBioisostere can address the compatibility of replacement with the whole molecular structure, which is necessary to optimize delicate molecular properties such as synthetic accessibility and drug-likeness.
Furthermore, we demonstrate the ability of our model to identify unexplored bioisosteric replacements that achieve the target property, which is beyond the capacity of conventional approaches.
% It also incorporates statistical patterns from experimental bioactivity data so that the proposed replacements preserve the molecule’s bioactivity.
Finally, we apply DeepBioisostere in an in silico hit-to-lead workflow and demonstrate its ability to improve key biochemical properties of hit candidate molecules while maintaining binding affinity, offering a new opportunity toward automated drug design.}

\section{Results}
\subsection{Brief Introduction of DeepBioisostere}
\label{sec:method-deepbioisostere-architecture}
In this work, we regard chemical modification based on bioisosterism as a sequential process comprising three essential steps: (1) removal fragment selection, (2) insertion fragment selection, and (3) attachment orientation prediction.
These three steps are sequentially applied to complete molecular optimization.
To achieve desirable structural modification, target biochemical properties dictated by the molecular structure should be considered during the three steps. For this purpose, we devised a deep generative model named DeepBioisostere to design a bioisosteric replacement for the multi-property control of a particular molecule.

The DeepBioisostere model is trained on an MMP dataset with structural analogs with minor differences, which we constructed from the ChEMBL database.
Since we designed the model with three main modules corresponding to each step in the chemical modification process, it learns the proper ways to remove and then insert fragments to meet the given conditions of multiple biochemical properties.
This sequential modification facilitates the optimization of a molecule while retaining its overall properties, except for the target ones to be improved. 
Notably, the DeepBioisostere model comprises two embedding networks at different levels: atom and fragment.
This unique feature, especially the latter network, enables the model to learn the compatibility between an insertion fragment and the remaining parts, allowing for fine control of complex properties that depend on the entire molecular structure.
For synthetically accessible chemical modification, the DeepBioisostere model enumerates molecular fragments within a given molecule based on Breaking Retrosynthetically Interesting Chemical Substructures (BRICS)\cite{BRICS} rules, and selects one of them as a removal fragment.
Then, the model selects an insertion fragment from a pre-defined fragment library by examining all of its items in terms of their compatibility with the remaining part of a given molecule and the target property control condition.
More details about the training and generation processes of the DeepBioisostere model can be found in \cref{sec:methods}.

\subsection{Multi-property Control with DeepBioisostere}
\label{sec:multi-property-control}

We first apply our DeepBioisostere model to multi-property control tasks.
During the optimization of a potent molecule in drug development, its structure is modified and elaborated to improve biochemical properties, such as lipophilicity, water solubility, synthetic accessibility, and drug-likeness.
This so-called multi-property control is more challenging than controlling a single property since, for example, the lipophilicity can be increased simply by elongating a carbon chain, but the modified structure is often not drug-like.\cite{2019optimization}
Thus, a comprehensive consideration of the whole ligand structure and its properties is required to modulate multiple properties simultaneously.

Here, the goal is to provide modified molecular structures to meet a given multi-property condition.
We tested our model for three scenarios that practitioners might encounter in a lead optimization process: (1) increasing or decreasing ligand lipophilicity while maintaining its molecular size, (2) increasing drug-likeness while maintaining its molecular weight, and (3) alleviating the synthetic complexity of a ligand with fair bioactivities.
The second and third scenarios are more challenging than the first, which requires more comprehensive consideration of the relationship between the properties of interest.
% Throughout \cref{sec:multi-property-control}, the following criteria are commonly employed to select test molecules: (1) a single neutral molecule, (2) absence of more than three rings, (3) inclusion of at least one BRICS bond, and (4) molecular weight that falls within the range of 250 to 500.
\edit{Throughout \cref{sec:multi-property-control}, the following criteria are commonly employed to select test molecules: (1) a single neutral molecule, (2) molecular weight that falls within the range of 100 to 500, and (3) inclusion of at least one BRICS bond.}
In addition, to examine our model's capability to find novel bioisosterism, we only used fragments from the test dataset for insertion, which were never adopted either as ground truths or negative samples for the insertion step in the training process.

The overall property control results are exhibited in \cref{table:multi_prop_control}.
For scenario 1, we selected 1,000 unique molecules whose log$P$ values are within the range of 1 to 5, regarded as moderate lipophilicity.
Whether one should increase a molecule's log$P$ or not depends on its specific application; for example, the biological target under consideration or the drug delivery plan are critical factors in making such a decision.
Thus, we tested our model for both cases with two property control conditions: increasing log$P$ by 1 and decreasing log${P}$ by 1 while maintaining overall molecule size in terms of molecular weight (MW).
We note that log$P$ is a logarithm of the octanol-water partition coefficient, so the difference by 1 corresponds to a 10 times larger or smaller solubility ratio.\cite{logp}
We generated {ten} unique molecular structures for each original molecule according to the two property control conditions.
For a few test molecules, DeepBioisostere generated fewer than {ten} molecular structures for each one since the BRICS-allowed insertion fragments were fewer than {ten}.
The results are summarized in the first row on \cref{table:multi_prop_control}.
We observed that the mean values of the molecular weight differences between an original molecule and its modified ones were almost negligible: +0.04 and -0.28, respectively.
These values suggest that the molecular weights of most modified structures are different from the original ones by less than one heavy atom. (See Supplementary Figure 4 for more details about property distributions)
Meanwhile, log$P$ was modulated with averaged changes of +0.83 and -0.84, falling slightly behind the requested property values.
These statistics indicate that our model successfully adjusted the lipophilicity of the given molecular structures with barely changing their molecular weights, which requires careful selection of both removal and insertion fragments.

\begin{table}[t!]
    \centering
    \resizebox{0.8\textwidth}{!}{
    {
    % \color{blue}
    \begin{tabular}{ l c c c c }
    
    \toprule

    \multirow{2.5}{*}{\textbf{Scenario}} & \multicolumn{2}{c}{\textbf{Requested Optimization}} & \multicolumn{2}{c}{\textbf{Property Change}} \\
    \cmidrule(lr){2-3} \cmidrule(lr){4-5}
    & Property &
    \multicolumn{1}{c}{$\Delta$Value} & 
    Average &
    Std. Dev. \\

    \midrule

    \multirow{4.5}{*}{Scenario 1} & MW& 0 & +0.04 & 12.8 \\
    & log$P$ & +1 & +0.83 & 0.38 \\
    
    \cmidrule{2-5}
    
    & MW& 0 & -0.28 & 12.6 \\
    & log$P$ & -1 & -0.84 & 0.43 \\
    
    \midrule
    
    \multirow{4.5}{*}{Scenario 2}
    % & $MW$ & 0 & +2.73 & 12.7 \\
    % & QED & +0.1 & +0.064 & 0.098 \\
    & MW& 0 & -0.11 & 12.1 \\
    & QED & +0.1 & +0.067 & 0.090 \\
    
    \cmidrule{2-5}
    
    % & $MW$ & 0 & +1.90 & 12.4 \\
    % & QED & +0.2 & +0.087 & 0.101 \\
    & MW& 0 & -0.57 & 12.0 \\
    & QED & +0.2 & +0.100 & 0.099 \\
    
    \midrule
    
    \multirow{4.5}{*}{Scenario 3}
    % & QED & 0 & -0.049 & 0.092 \\
    % & SAscore & -0.5 & -0.294 & 0.242 \\
    & QED & 0 & -0.026 & 0.073 \\
    & SAscore & -0.5 & -0.424 & 0.433 \\
    
    \cmidrule{2-5}
    
    % & QED & 0 & -0.050 & 0.09 \\
    % & SAscore & -1.0 & -0.501 & 0.224 \\
    & QED & 0 & -0.028 & 0.074 \\
    & SAscore & -1.0 & -0.719 & 0.524 \\

    \bottomrule
    
    \end{tabular}}
    }
    \caption{\textbf{Results of three multi-property control scenarios.} The requested optimization is the property control condition given to our model. For each scenario, we have two properties to control: Scenario 1 - modulating log$P$ while maintaining MW. Scenario 2 - increasing QED while keeping the molecular weight. Scenario 3 - decreasing SAscore without losing QED. $\Delta\textnormal{Value}$ is the requested property change in each property. $\textnormal{Average}$ and $\textnormal{Std. Dev.}$ designates the mean value of property change and its standard deviation, respectively.} 
    \label{table:multi_prop_control}
\end{table}

\begin{figure*}[ht!]
 \centering
 \includegraphics[width=0.87\textwidth]{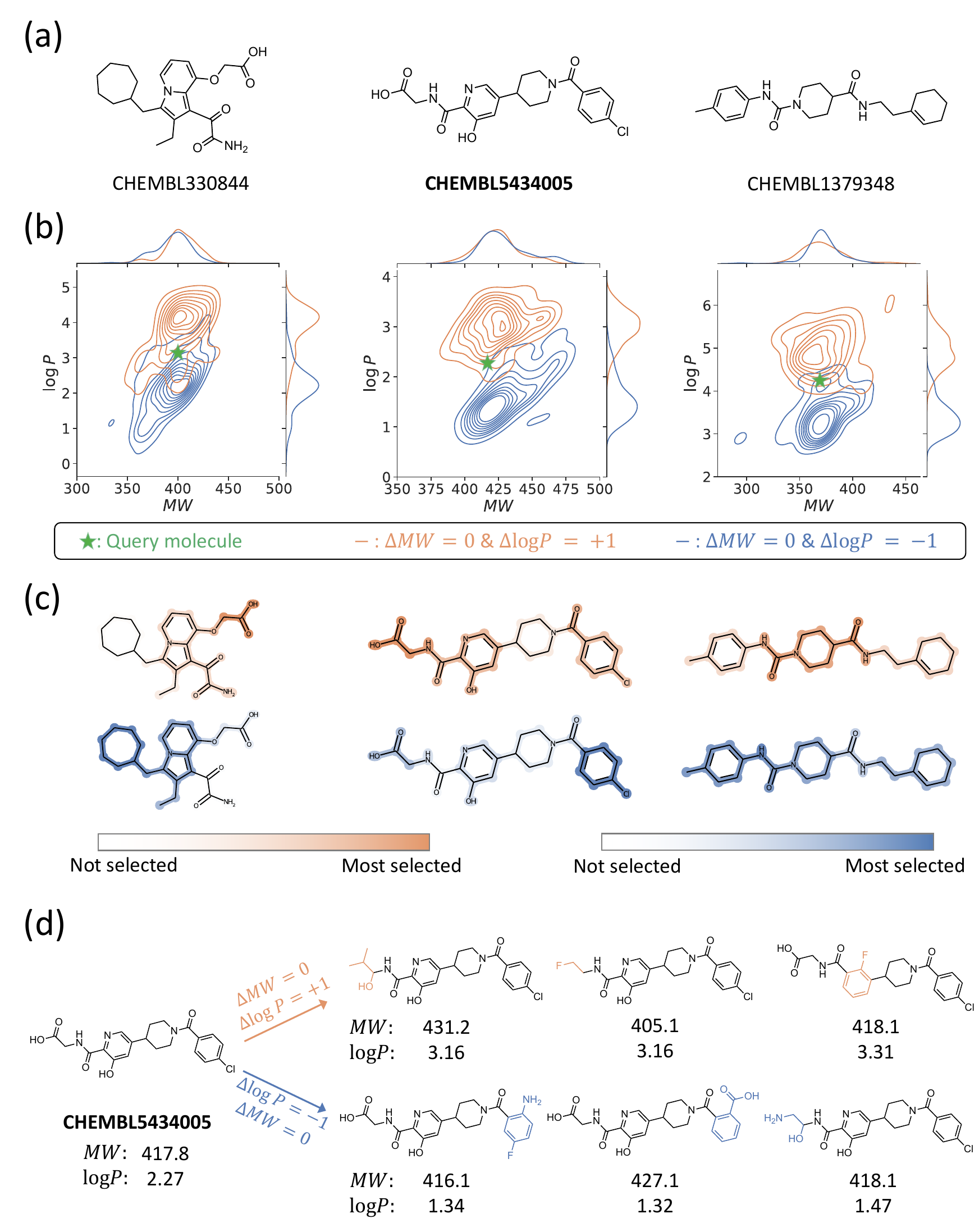}
 \caption{\textbf{Case studies on the multi-property control with DeepBioisostere for scenario 1, described in \cref{table:multi_prop_control}.} In this figure, orange-colored ones are results for increasing log$P$ by +1, and blue-colored ones are for decreasing log$P$ by -1, both with maintaining molecular weights. (a) Three original molecules are to be modified. (b) Two-dimensional distribution plots of 100 unique modified molecules for each original molecule. (c) Removal fragment selection results. The thicker a fragment is shaded, the more it was chosen as a removal fragment for property control. The upper orange-colored ones visualize the results for increasing log$P$, and the lower blue-colored ones visualize the results for decreasing log$P$. (d) Two series of the top-ranked structures were derived from the second test molecule, CHEMBL5434005.}
 \label{figure:figure1. multi-conditioning}
\end{figure*}

Second, we demonstrated an optimization process where a molecular structure is modified to improve its drug-likeness in terms of quantitative estimation of drug-likeness (QED).\cite{qed} (Scenario 2 in \cref{table:multi_prop_control})
In this scenario, only molecules whose QED values are less than 0.4 were used as test molecules.
Property control conditions were set to increase the QED value by 0.1 or 0.2 while keeping the molecular weight constant.
By employing the DeepBioisostere model, we were able to obtain novel molecules with enhanced drug-likeness; the mean values of QED changes were +0.067 and +0.100, respectively.
This result indicates that our model can modulate drug-likeness without much change in the original molecular structure, which is critical in order not to lose pre-established crucial properties.
Although the property control condition specifying a larger QED increase yielded molecules with higher QED values, the deviation from the given condition also became larger.
It is not surprising that increasing QED by 0.2 is even more challenging, as the occurrence of such replacements within our training dataset is exceedingly rare.
Refer to supplementary Section 4.1 for further discussion about the statistics.

Lastly, we tested our model to enhance the synthetic accessibility of a molecule with a high QED value but low synthetic accessibility. (scenario 3 in \cref{table:multi_prop_control})
Here, we adopted the SAscore introduced by \citet{sascore}, a widely used estimator of synthetic accessibility.
We collected test molecules for this scenario with the criteria that the QED value is greater than 0.7 and the SAscore is greater than 4.
Note that the SAscore of a simple catalog molecule is about 2 or 3, whereas for a natural product, it is mostly greater than 5.\cite{sascore}
Property control conditions were set to decrease the SAscores of molecules by 0.5 or 1.0 while retaining their QED values.
In this scenario, our model provided novel molecules with lower SAscores, although several of them exhibited somewhat decreased QED values.
This result suggests that DeepBioisostere can be adopted to supplement many in silico methods for drug design, whose outcomes are often questioned regarding their synthetic accessibility.\cite{walters2018virtual, gao2020synthesizability}

\subsection{Detailed Analysis on Modification with DeepBioisostere}
\label{sec:analysis-modification-with-deepbioisostere}
In \cref{sec:multi-property-control}, we demonstrated that DeepBioisotere can control multiple molecular properties simultaneously while modifying molecular structures.
Next, we analyze the modification process in detail in terms of two major steps: the removal fragment selection and the insertion fragment selection.
Careful decisions in these steps are essential to control molecular properties, where the overall molecular structure is to be determined.
Here, we carry out case studies for multi-property control and will discuss the removal fragment selection and the insertion fragment selection sequentially.

We randomly selected three molecules from the ChEMBL database with moderate molecular weight and lipophilicity, as visualized in \cref{figure:figure1. multi-conditioning}(a).
We suppose scenario 1 introduced in \cref{sec:multi-property-control}, where log$P$ is to be modulated while a molecular weight is kept unchanged.
We generated 100 unique modified molecular structures for each test molecule and property condition with the DeepBioisostere model.
Again, only the fragments from the test dataset for insertion were adopted, the same as in the previous section.
We exhibited their log$P$ values and molecular weights in \cref{figure:figure1. multi-conditioning}(b).
For every test molecule, the generated structures under different property conditions showed distributions that were completely distinct in terms of log$P$ values but very similar in molecular weights.
log$P$ was increased or decreased by about 1 while most changes in molecular weight were less than 20, corresponding to the weight of one or two heavy atoms.

\subsubsection{Removal Fragment Selection}
\label{sec:removal-fragment-selction}
We looked into the generated molecules and analyzed the removal fragment selection step. We counted how many times each substructure was selected as a removal fragment and visualized them in \cref{figure:figure1. multi-conditioning}(c).
For the molecule in the middle column, CHEMBL5434005, the carboxylic acid group was most frequently selected as a removal fragment to increase log$P$.
In contrast, to decrease log$P$, it was hardly substituted, but the chlorobenzene moiety was preferred as a removal fragment.
These choices of removal fragments align well with the known chemical knowledge; introducing a carboxylic acid group enhances hydrophilicity, whereas incorporating an aryl halide moiety increases hydrophobicity.
The other two test molecules also showed similar differences in the removal fragment selection under different property control conditions, as exhibited in the 1st and 3rd columns in \cref{figure:figure1. multi-conditioning}.

% For the second test molecule in \cref{figure:figure1. multi-conditioning}(a), CHEMBL1522368, we visualized top-3 ranked structures generated by DeepBioisostere in \cref{figure:figure1. multi-conditioning}(d).
We further visualized three analogs for CHEMBL5434005 generated by DeepBioisostere with the highest likelihoods in \cref{figure:figure1. multi-conditioning}(d).
% These examples illustrate the model's ability to generate diverse molecules by suggesting specific bioisosteric replacements to modulate log$P$ while aiming to maintain the molecular weight relatively constant ($\Delta MW\approx 0$).
% When required to increase the log$P$ value (the orange arrow), our model substituted a nitrofuran ring with another fragment.
% All the selected three insertion fragments are aryl groups with halogen elements, which are apparently more hydrophobic than the nitrofuran ring.
% When conditioned to increase log$P$ (the orange arrow), DeepBioisostere proposed modifications to decrease polarity and enhance lipophilicity.
When conditioned to increase log$P$ (orange arrow), the carboxylic acid group of the parent molecule was replaced with an isopropanol or fluoroethyl group, and the hydroxypyridine scaffold was substituted by a fluorobenzene ring, which are the cases where bioisosteric replacements that introduce more polarity were applied.
As a result, the proposed molecular structures exhibited log$P$ values higher than the original.
% Meanwhile, when required to decrease the log$P$ value (the blue arrow), our model selected a 2,5-dimethyl phenyl group to replace with other aryl substituents.
% The alternative fragments contain a hydroxyl group or a pyridine moiety, resulting in the decreased log$P$ values of the modified structures: 1.34, 1.32, and 1.47.
Meanwhile, to reduce the log$P$ value (blue arrow), our model selected the 4-chlorophenyl moiety on the terminal site of CHEMBL5434005 as a key site for modification.
The chosen alternative fragments introduce more hydrophilic aryl groups, reducing the modified structures’ log$P$ values.
In the last example, the DeepBioisostere model selected a carboxylic acid group, despite its well-known hydrophilicity, as the removal fragment.
We believe this is due to the fact that a carboxylic acid is one of the most frequently considered bioisosteres in the hit-to-lead process, which should be reflected in the training data we constructed.\cite{lassalas2016structure}
Importantly, our model still succeeded in lowering the log$P$ of the given molecule by introducing a more hydrophilic moiety as an alternative.
These results suggest that our model's sequential-decision framework discussed in \cref{sec:method-deepbioisostere-architecture} effectively balances property-control objectives by choosing appropriate removal fragments and selecting insertion fragments conditioned on diverse combinations.

\subsubsection{Insertion Fragment Selection - Effect of Chemical Environment on Property Control}
\label{sec:insertion-fragment-selection}
Another essential factor that governs the selection of the insertion fragments is the chemical environment.
Here, we refer to the chemical environment as the substituents surrounding the selected removal substructure to be replaced by a new insertion fragment.
We demonstrate its importance in molecular optimization by two single-property control tasks: increasing log$P$ by 1 and increasing QED by 0.1.
% Obviously, the latter task is more complicated to achieve than the former since QED is calculated based on eight individual molecular features, including log$P$.

\begin{figure}[t!]
 \centering
 \includegraphics[width=1.0\textwidth]{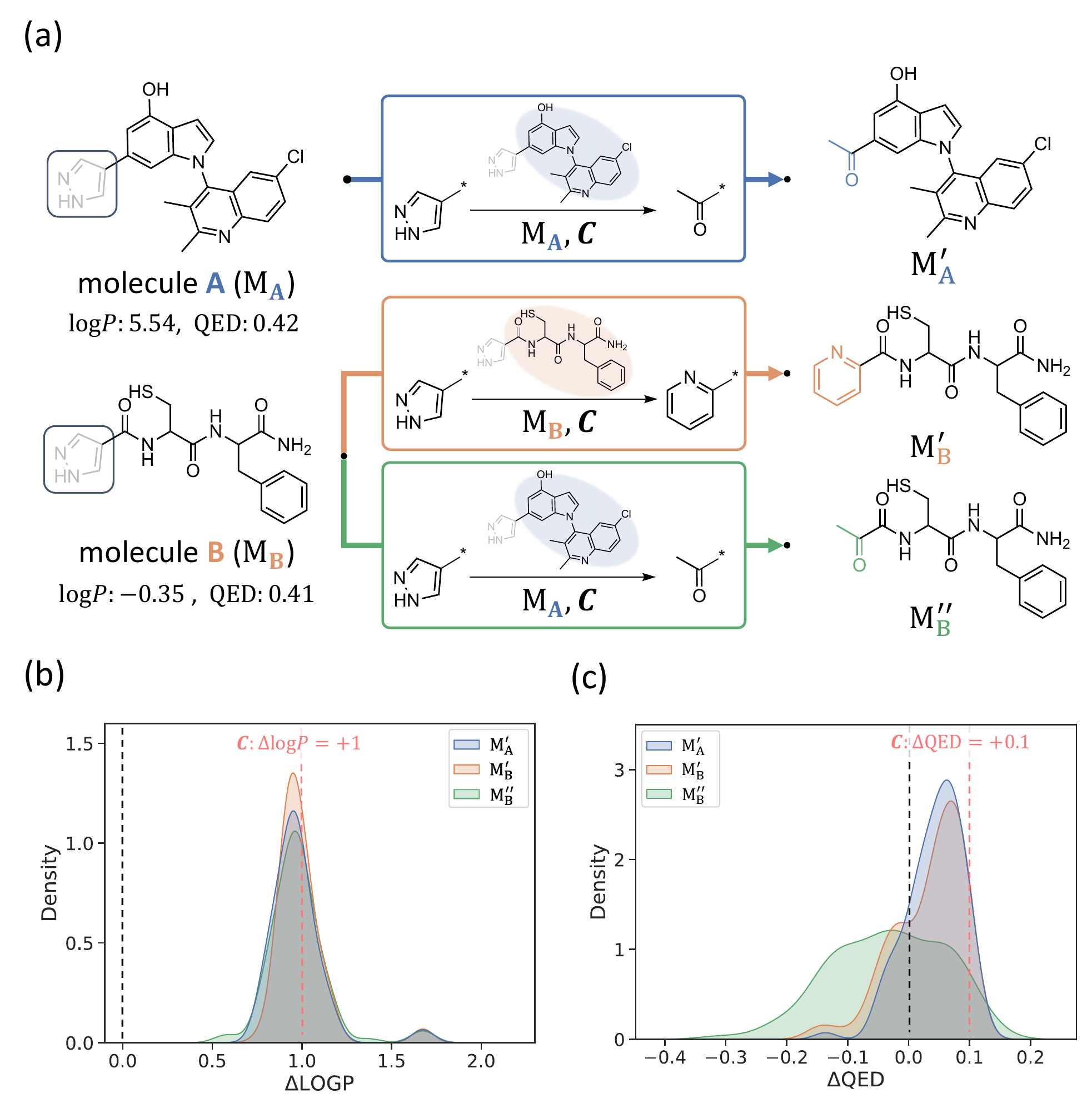}
 \caption{\edit{\textbf{An example showing the significance of chemical environment during insertion fragment selection.} (a) Three replacement cases: within molecule A ($M'_A$), within molecule B ($M'_B$), and from molecule A to molecule B ($M''_B$). (b) $\Delta \log{P}$ distributions of modified molecules from each original molecule. (c) $\Delta$QED distributions of modified molecules for each original molecule. In both (b) and (c), the black and red dashed lines denote the baseline implying original molecules' property values and property conditioning objectives, respectively.}}
 \label{figure:effect-of-chemical-environment}
\end{figure}

\cref{figure:effect-of-chemical-environment} illustrates the overall experimental procedure and its results.
First, we selected two molecules that contain a pyrazole group, which is a frequently considered aromatic group in bioisosteric replacement.\cite{phenyl-bioisosteres}
Their molecular structures are shown in \cref{figure:effect-of-chemical-environment}(a); we refer to them as molecules A and B (or $M_A$ and $M_B$, respectively).
The two molecules exhibit great differences in log$P$(5.54 and -0.35) while showing moderate levels of QED values(0.42 and 0.41).
Thus, the pyrazole moieties in the two molecules are situated in fundamentally different chemical environments, particularly in terms of the number of aromatic rings, hydrogen bond donors and acceptors, and rotatable bonds.
Here, we forced our model to choose only the pyrazole groups as removal fragments to modulate a given target property, resulting in 100 unique modified structures. (\textcolor{blue}{$M'_A$ and $M'_B$ in \cref{figure:effect-of-chemical-environment}(a)})
For comparison, we retrieved the insertion fragments selected for molecule A and incorporated them into molecule B, resulting in another set of 100 unique molecules. (\textcolor{blue}{$M''_B$} in \cref{figure:effect-of-chemical-environment}(a))
In the first two cases, \edit{$M'_A$} and \edit{$M'_B$}, the individual chemical environments were directly engaged in the selection of insertion fragments.
However, in the case of \edit{$M''_B$}, the insertion fragments were selected based solely on the chemical environment of molecule A, without any involvement from molecule B.
By comparing these cases, we can examine the significance of the chemical environment in molecular property control, especially during the selection of insertion fragments.

\cref{figure:effect-of-chemical-environment}(b) and (c) show the property distributions of the modified molecules, where both \edit{$M'_A$} and \edit{$M'_B$} nearly achieve the objective values.
However, in the case of \edit{$M''_B$}, the observed property changes in log$P$ and QED conditioning tasks were distinct.
In the former task, insertion fragments selected to increase log$P$ of molecule A also increased log$P$ of molecule B, despite the differences in the chemical environments in which the pyrazole moieties were involved.
Conversely, in the QED conditioning task, the case of \edit{$M''_B$} resulted in molecular structures with decreased QED values, while those from \edit{$M'_A$ and $M'_B$} exhibited increased QED values compared to the original ones.
This apparently shows that the bioisosteric replacements selected to improve the QED value of molecule A failed to improve that of molecule B.
We attribute this distinction to the intrinsic difference between the two properties; log$P$ can be readily approximated by summing over the estimated contributions of each chemical moiety in a molecule, while QED is a non-linear function of the molecule-level structural features, such as the numbers of hydrogen bond donors and acceptors.
% This example study suggests that the surrounding parts, not only the removal fragment, should be considered when choosing a suitable bioisosteric replacement to modulate a target molecular property dependent on the whole molecular structure in a broad sense. 
% This is especially important in drug design, where crucial properties such as synthetic accessibility and ligand binding affinity depend on the non-linear relationships between internal substructures or structural features.
\edit{This example highlights the importance of considering the entire chemical environment, not just the removed fragment, when designing a suitable bioisosteric replacement for properties that depend on global molecular context.
Such context-dependent property modulation cannot be handled by database-mining approaches, which rely solely on pre-defined fragment statistics.
In contrast, DeepBioisostere explicitly learns to capture the non-linear relationships between fragments and their molecular environments, enabling context-aware bioisosteric replacements.}

\begin{figure}[t!]
 \centering
 \includegraphics[width=1.0\textwidth]{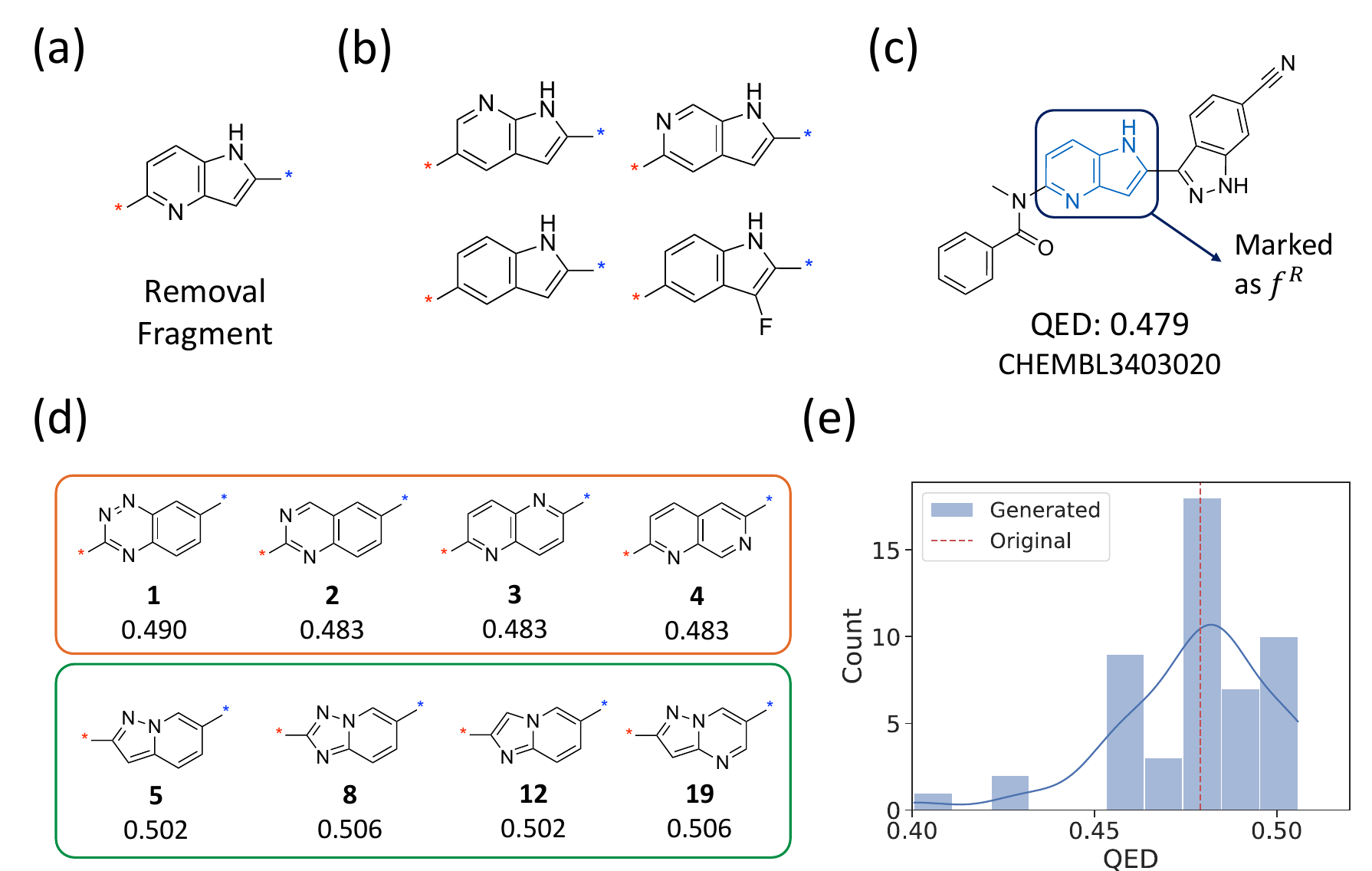}
 \caption{\textbf{An example to find bioisosteres of an uncommon molecular fragment.} (a) A 4-azaindole moiety, an example of an uncommon moiety. (b) Bioisosteres of the 4-azaindole linker identified from our MMP database. (c) A test molecule from the ChEMBL database, which was modified by our model with its azaindole moiety marked as a removal fragment. (d) Eight top-ranked fragments proposed by our model as bioiosteres. The fragments fall into mainly two types, grouped by different colored boxes for each type. The integer placed below each fragment indicates the order ranked by the DeepBioisostere model, and the decimals are the QED values of the resultant molecules. In (a), (c), and (d), the red and blue stars (*) indicate the corresponding connection sites, respectively. (e) The QED distribution of the 50 generated molecules. The red dashed line denotes the QED value of the original molecule.}
 \label{figure:figure3. novel bioisosterism}
\end{figure}

\subsection{Exploring Bioisosteres for Uncommon Molecular Fragments}
\label{sec:finding-novel-bioisosterism}
Next, we examine our model in identifying potential bioisosterism.
Identification of novel bioisosteres becomes necessary when modification of an uncommon molecular fragment is required.
In this case, a typical database-mining approach is likely to be insufficient since only a few bioisosteres from the database might be available for that fragment.
Indeed, among \edit{140,193} fragments in the fragment library we created, \edit{98,459} fragments are found to have no more than ten bioisosteres, respectively.
\cref{figure:figure3. novel bioisosterism} illustrates such an example of modifying an uncommon fragment.
The molecular fragment in \cref{figure:figure3. novel bioisosterism}(a), a 4-azaindole moiety, appears dozens of times in the ChEMBL database as a linker in the visualized fashion.
Utilizing our MMP database, only four fragments were found as bioisosteres, and we visualized them in \cref{figure:figure3. novel bioisosterism}(b).
All the found bioisosteres are fused aromatic heterocycles, the same as the original fragment except for one heavy atom, which can form similar non-covalent interactions with biological targets.
% However, by solely depending on these fragments identified from the MMP analysis, one cannot explore diverse chemical moieties to find potential bioisosteres.
However, by solely depending on these fragments identified from the MMPA, one cannot explore diverse structural analogues of an acclaimed molecule.
More statistics about the number of bioisosteres for each fragment in the fragment library can be found in Supplementary Figure 3.

To find more bioisosteres beyond them, we employed our model on a molecule from the ChEMBL database by specifying the azaindole moiety as a removal fragment, as in \cref{figure:figure3. novel bioisosterism}(c).
We allowed all the insertion fragments in our fragment library regardless of whether they were used in the training process.
Here, to explore bioisosteres of the azaindole moiety that exhibit similar drug-likeness, we set the property control objective as $\Delta$QED=0.
With this setting, we generated 50 unique molecular structures from the original molecule with DeepBioisostere.

\cref{figure:figure3. novel bioisosterism}(d) illustrates eight insertion fragments predicted by our model.
% We noticed that they fall into mainly two types: fused aromatic rings (the orange box) and covalently bonded aromatic rings (the green box).
% The fragments in the orange box consist of fused aromatic heterocycles, each containing one or two nitrogen atoms, similar to those identified from our MMP database shown in \cref{figure:figure3. novel bioisosterism}(b).
% On the other hand, the fragments in the green box have substantially different molecular structures, where each of them contains covalently bonded aromatic rings.
Under the constraint of $\Delta$QED=0, several fused aromatic rings containing a few hetero-atoms were chosen as bioisosteres of the azaindole moiety.
% They can be divided into two types: composed of two six-membered rings (the orange box) and one six-membered ring and one five-membered ring (the green box).
These fragments feature either a six-membered ring fused to a five-membered ring (the green box) or two six-membered rings, in contrast to the original azaindole moiety (the orange box).
Despite this variation, all highly ranked fragments, when inserted into the test molecule, exhibit QED values comparable to the original. (see \cref{figure:figure3. novel bioisosterism}(d))
Each fragment retains a similar number of hydrogen bonding sites, two aromatic rings, and comparable lipophilicity, suggesting that they can engage in analogous non-covalent interactions.
\cref{figure:figure3. novel bioisosterism}(e) illustrates the QED distribution of 50 unique modified structures.
The peak near the original QED value implies that most insertions retain key biochemical features, including hydrogen-bond capacity, aromaticity, and polar surface area.
% This example study demonstrates that one can adopt DeepBioisostere for an uncommon chemical moiety to explore potential bioisosteres with a similar level of drug-likeness or QED.
This case study demonstrates that DeepBioisostere can be effectively applied to identify potential bioisosteres of uncommon chemical moieties that exhibit drug-likeness profiles comparable to those of the parent moiety.

% original subsection title: Optimizing Computational Hit Candidates Derived from Generative Models
\subsection{\edit{In Silico Hit-to-lead Optimization of Structure-based Generative Model-derived Hit Candidates}}
\label{sec:optimizing_generative_models}
We further extend the scope of our model to an in silico hit-to-lead workflow, where model-generated hit candidates are refined to improve drug-likeness and synthetic accessibility. 
%optimize computational hit candidates, generated by structure-based 3D generative models.
Recent 3D molecular generative models for structure-based drug design have focused more on binding affinity to a given target protein, especially considering the structural context of the pocket environment in which the interaction between a protein and a generated molecule occurs.\cite{pocket2mol,deepicl,targetdiff,decompdiff}
Despite the target-specific design capacity of these approaches, generated molecules often fail to meet the criteria of experimental hits due to other molecular properties. 
%This is due to the low synthesizability or drug-likeness, which are the molecular properties that rely closely on the chemical structure of the molecule rather than its 3D conformation.

%Recently emerging 3D molecular generative models primarily focus on creating molecules with high binding affinities to specific protein targets.
%By leveraging detailed 3D structural information of protein-ligand interactions, these models can explore vast chemical spaces to design molecules with complementary pharmacophoric features.
%However, while optimizing binding affinity is advantageous, a singular focus on this aspect often results in molecular candidates impractical for clinical progression.
%For instance, characteristics such as SAscore or QED are frequently underexplored by these 3D-centric approaches, which rely less heavily on 2D topological and physicochemical descriptors.
%presents a significant challenge, as potent binders may not be viable drug candidates owing to these deficiencies, 
This gap between computational and actual hits highlights the need for a robust strategy to refine model-generated hit candidates, or in other words, to enhance molecular properties while preserving binding affinities.
Hence, we demonstrate the synergy of using DeepBioisostere for multi-property optimization on top of target-specifically designed molecules from molecular generative models, revealing the applicability of our strategy.
%From a practical perspective, our strategy mirrors the refinement of hit candidates to yield potential lead compounds.}
%from a more practical perspective through a case study that mimics the crucial hit-to-lead process, specifically by refining such computationally derived starting points.
The goal here is to optimize synthesizability and drug-likeness of hit candidates, estimated as SAscore and QED, respectively, without losing their bioactivities towards respective targets.

We initially obtained computational hit candidates by four recent state-of-the-art structure-based 3D molecular generative models: Pocket2Mol\cite{pocket2mol}, DeepICL\cite{deepicl}, TargetDiff\cite{targetdiff}, and DecompDiff\cite{decompdiff}.
% We selected three target proteins from the CrossDocked2020 test set based on average QED and SAscores of the pre-generated molecules of each target.
%From the CrossDocked2020 test set, we selected three target proteins whose generated molecules exhibited the poorest average QED (lowest) and the least favorable SAscores (highest).
From the CrossDocked2020 test set, we selected three target proteins whose generated molecules exhibit both poor drug-likeness and synthetic accessibility, making them suitable targets that can take the most advantage from chemical modification.
% suitable targets for the starting point of chemical modification.
The details of the model used and the criteria for target selection, along with the statistics of the generated molecules, are provided in the Supplementary Information Section 7.
% We note that the three former baselines share the same fragment replacement candidates; the only difference is their ranking strategy of given replacement candidates.
Generated molecules for each target and model, termed computational hit candidates, underwent modifications through different replacement selection strategies, including DeepBioisostere.
For comparison of our model with conventional database-mining methods, we adopted three baseline replacement strategies, which are referred to as random, frequency-based, and MMPA-based, as shown in \Cref{tab:generative_model_summary}.
The three strategies were set to utilize the bioisosteric pairs in the MMP database we constructed.
In particular, the MMPA-based strategy first filters out replacements whose occurrence in our database is less than 10 times, and then selects replacements that exhibited the highest improvement in both QED and SAscore in the database.
This process is based on statistical evidence in the MMP database, in the same way with the conventional database-mining strategies, such as SwissBioisostere.
Here, we guide DeepBioisostere to increase QED by 0.1 and to decrease SAscore by 1.0, simultaneously.
For each method, replacement was attempted 100 times for each input molecule.

\begin{table}[ht!]
    \centering
    % \resizebox{\textwidth}{!}{%
    \begin{tabular}{llccc|c}
        \toprule
        \multirow{2.5}{*}{\textbf{Model}} & \multirow{2.5}{*}{\textbf{Strategy}} & \multicolumn{4}{c}{\textbf{Success Rate}} \\
        \cmidrule{3-6}
         & & \textbf{QED} & \textbf{SAscore} & \textbf{Docking} & \textbf{Joint} \\
        \midrule
        \multirow{4.5}{*}{DeepICL} & Random & 0.19 & 0.17 & 0.97 & 0.06 \\
                                 & Frequency-based & 0.30 & 0.26 & \bfseries 0.99 & 0.10 \\
                                 & MMPA-based & 0.33 & 0.22 & 0.97 & 0.11 \\
                                    \cmidrule{2-6}
                                 & DeepBioisostere & \bfseries 0.69 & \bfseries  0.80 & 0.96 &\bfseries 0.56 \\
        \midrule
        \multirow{4.5}{*}{Pocket2Mol}& Random & 0.62 & 0.39 & 0.91 & 0.24 \\
                                    & Frequency-based & 0.67 & 0.51 & 0.97 & 0.36 \\
                                    & MMPA-based & 0.64 & 0.42 & \bfseries 0.98 & 0.30 \\
                                    \cmidrule{2-6}
                                    & DeepBioisostere & \bfseries 0.75 & \bfseries 0.74 & 0.94 &\bfseries 0.56 \\
        \midrule
        \multirow{4.5}{*}{TargetDiff}& Random & 0.32 & 0.50 & 0.96 & 0.18 \\
                                    & Frequency-based & 0.43 & 0.58 & 0.98 & 0.25 \\
                                    & MMPA-based & 0.42 & 0.50 & \bfseries 0.99 & 0.20 \\
                                    \cmidrule{2-6}
                                    & DeepBioisostere & \bfseries 0.83 & \bfseries  0.87 & 0.97 &\bfseries 0.74 \\
        \midrule
        \multirow{4.5}{*}{DecompDiff} & Random & 0.38 & 0.39 & 0.95 & 0.19 \\
                                 & Frequency-based & 0.49 & 0.45 & 0.98 & 0.24 \\
                                 & MMPA-based & 0.43 & 0.29 & \bfseries 0.99 & 0.13 \\
                                    \cmidrule{2-6}
                                 & DeepBioisostere & \bfseries 0.79 & \bfseries  0.87 & 0.96 &\bfseries  0.68 \\
        \bottomrule
    \end{tabular}
    \caption{
    \textbf{Success rate of in silico hit-to-lead workflow.}
    Starting from the model-derived hit candidates, four different replacement strategies are applied to achieve the goal of multi-property control---enhancing QED and SAscore while maintaining the docking score. The best success rate among strategies is highlighted in bold.
    % target id 68
    }
    \label{tab:generative_model_summary}
\end{table}

For evaluation, we compared the success rates of QED and SAscore optimization, as well as the retention of docking scores to a protein target.
% These metrics are based on the objective of this experiment, which is to enhance the drug-likeness and synthetic accessibility of candidate molecules while maintaining their binding affinity to the protein target.
The success criteria for QED and SAscore are based on whether the generated molecule exhibits better properties than the original.
%The criteria for property optimization were based on improvements, meaning the generated molecule exhibited better properties than the original.
We consider a generated docking score as a success when it is retained during replacement within 1.36 kcal/mol, corresponding to a 10-fold change in IC50.
%Docking score retention was defined as a difference of less than 1.36 kcal/mol between the generated and original molecules, corresponding to a 10-fold change in IC50.
The most integral metric, which is termed `Joint', means the ratio of generated molecules jointly meets all the aforementioned criteria---improved QED, SAscore, and preserved docking score---thereby indicating the success in the computational hit-to-lead scenario for increasing synthesizability and drug-likeness.
% To evaluate the generative efficiency of each method, we calculated a `generation ratio', defined as the number of successfully generated molecules divided by the total number of attempts.

\Cref{tab:generative_model_summary} shows the result of a target protein with PDB ID 4RV4, whereas the other results are shown in the Supplementary Table 4.
% Interestingly, rank-filtered MMPA, which we considered as a traditional bioisosteric replacement selection method, shows relatively better performance compared to random- and frequency-based methods.
Although frequency- and MMPA-based strategies performed well at docking score retention---possibly as a result of the use of our bioisosteric replacement library curated from bio-assays---DeepBioisostere mostly achieved superior performance in optimizing the properties such as QED and SAscore.
We attribute the improved controllability of QED and SAscore to the consideration of the entire molecular graph, along with the removal and insertion fragments, rather than focusing solely on a single fragment as in the baselines.
The true power of DeepBioisostere is demonstrated in its ability to satisfy all criteria simultaneously, as evidenced by its superior performance in the `Joint' metric, for which baseline approaches exhibited less than 10 \% success.

\begin{figure}[t!]
  \centering
    \includegraphics[width=\textwidth]{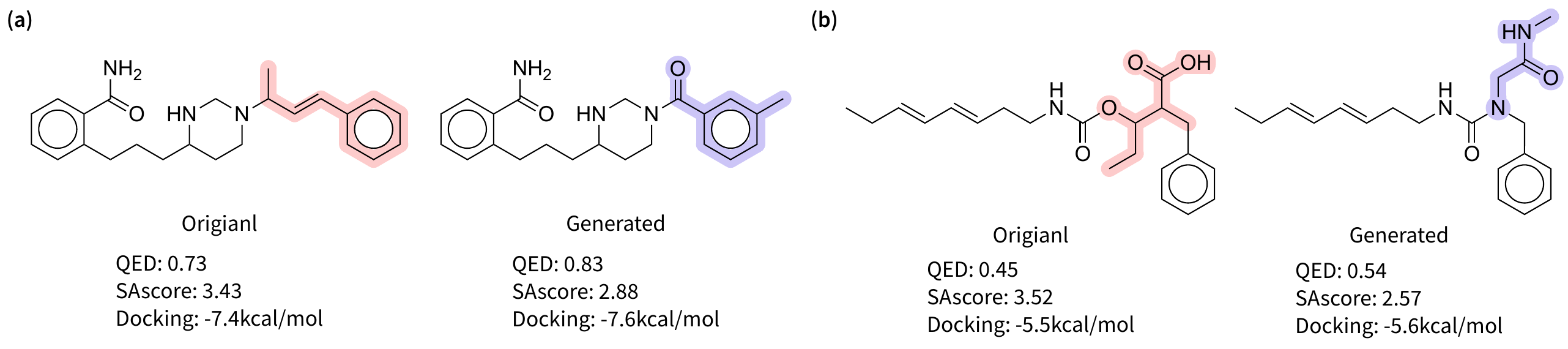}
  \caption{
  \textbf{Examples of generated molecules by DeepBioisostere for \cref{tab:generative_model_summary}.}
  The original molecules are selected from the generation results of DeepICL\cite{deepicl} on target 4RV4.
  }
  \label{fig:case_study_examples}
\end{figure}

\Cref{fig:case_study_examples} shows two example molecules generated based on bioisosteric replacements identified by DeepBioisostere, both of which achieve improved QED and SAscore while preserving the binding affinity of the original molecule.
Notably, these improvements arise from the substitution of the fragments resulting in amide-containing molecule , which is expected to enhance the synthetic accessibility.
Together, this robust ability to find replacements that fulfill multiple objectives is notably agnostic to underlying generative models, establishing it as a broadly applicable and powerful strategy for in silico drug design.

% 강조하고 싶은건
% 1. Frequency-based나 knowledge-based 보다 deepbioisostere가 multi-property objective를 만족하는 적절한 replacement를 잘 찾는다. (main)
% 2. 생성 모델에 agnostic하게 잘 적용된다. (minor)

%This selection spanned up to three distinct protein targets to assess the robustness of our methodology.
%For each protein target, a curated collection of 100 hit candidates generated by 3D molecular generative models was designated as the initial input for refinement by DeepBioisostere.

%Each of these 100 seed molecules per target was then individually processed using DeepBioisostere, with the objective of generating a diverse set of 100 new analogues for each input.
%The generation of these analogues was explicitly guided by multi-property optimization goals: increasing QED by 0.2 and decreasing SAscore by 1.0 to enhance the drug-likeness and synthesizability, respectively.
%Crucially, this bioisosteric replacement strategy inherently seeks to make minimal structural changes that are compatible with preserving the interaction between the protein and the original compound.
%Thus, we evaluated the ability of the model to improve multiple properties of given molecules without loss of binding affinity by considering all properties, including QED, SAscore, and binding affinity, calculated with Vina-GPU-2.1 at once.

\section{\edit{Discussion}}
\label{sec:conclusion}
In this work, we propose a deep generative model named DeepBioisostere for autonomous bioisosteric replacement with fine control over multiple molecular properties.
Our model conducts chemical modification in an end-to-end manner, from selecting fragments to remove and insert, along with their attachment orientation.
We demonstrate various scenarios of dual property control, where one property is altered while the other remains unchanged.
DeepBioisostere successfully modulated the properties of the given molecules in all scenarios by replacing the fragments in the desired manner.
We further demonstrate the adjustability of DeepBioisostere, which dynamically selects the fragment to remove based on the given property control condition.
This is practically advantageous, where previous database-mining approaches heavily rely on the help of experts for choosing a substructure to be modified.
Moreover, we justify the significance of substituents near the replacement site in controlling comprehensive molecular properties, especially for the selection of insertion fragments, which could not be addressed by conventional in silico approaches.
%Our model can design novel bioisosteric replacements that are not included in the training data by allowing flexible selections of fragment pairs. 
Our model enables more diverse replacements appropriate for the given property condition, extending from the existing moiety pairs within the database. 
We emphasize that this is particularly important when the removal fragment is uncommon, as it provides a wide variety of potential chemical modifications.
Finally, we applied DeepBioisostere to a practical scenario that mimics the hit-to-lead process, which involves optimizing the drug-likeness and synthetic accessibility of computationally generated hit compounds while retaining their binding affinities.
These results demonstrated that DeepBioisostere successfully tuned multiple properties without disrupting binding affinity for all 3D molecular generative models used, showcasing the promising role of our approach in autonomous drug design.

\edit{
While DeepBioisostere already demonstrates the ability to control multiple molecular properties through a deep generative framework, its design also opens promising directions for further enhancement. 
In particular, extending the model with multi-task and transfer learning could enable it to handle conditions with limited or noisy labels more effectively. 
Pretraining on widely available or easy-to-label molecular properties, such as QED, logP, and SA score, and transferring the learned representations to tasks involving complex endpoints, like ADMET properties, would enable the framework to generalize across a broader chemical and biological space. 
Such advancements could further strengthen the model’s applicability to diverse drug discovery pipelines, facilitating more accurate and data-efficient molecular design in increasingly realistic and challenging scenarios.
%While DeepBioisostere offers the advantage of enabling multi-property control --- including binding affinity --- through a deep generative framework, it also has still limitations.
%In particular, while our model can modulate molecular properties defined as control conditions, it shares a common limitation of conditional generative models: it can only effectively modulate properties with abundant and reliable labels.
%A promising direction to address this challenge is to leverage multi-task and transfer learning.
%By pretraining the model on properties with abundant labels (e.g., QED, logP, SA score) and transferring the learned molecular representations, the framework can better accommodate conditions with limited labels, such as specific ADMET properties.
%Such advancements could broaden the applicability of DeepBioisostere to a wider range of drug discovery tasks and more challenging molecular design scenarios.
}

% We could obtain potentially improved molecules that strongly interact with the mutant residue while retaining drug-likeness, synthetic accessibility, and size, showcasing the strength of our model.}

\section{Methods}
\label{sec:methods}

\begin{figure*}[ht!]
 \centering
 \includegraphics[width=1.0\textwidth]{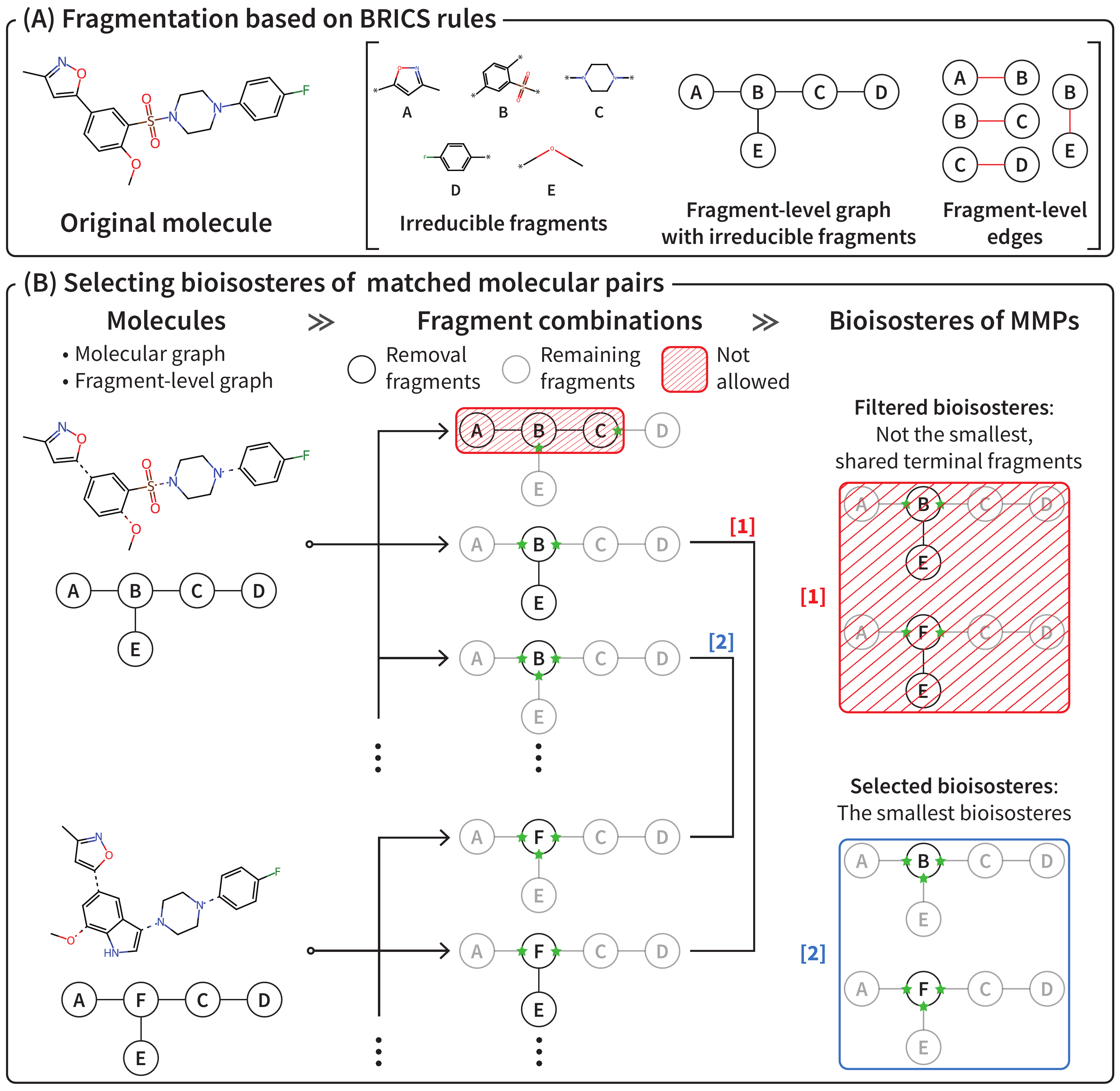}
 \caption{
 \textbf{The overall dataset generation scheme.}
 (a) The original molecule is fragmented according to the BRICS rules to derive irreducible fragments, resulting in a fragment-level graph with the corresponding edges, where each edge corresponds to a single BRICS bond.
 (b) The dataset is constructed by identifying bioisosteres.
 First, allowed fragment combinations are extracted for each molecule.
 Then, the smallest bioisosteres are selected for each matched pair of molecules that share the remaining fragments.
 The dataset incorporates all selected bioisosteres, along with the original molecule, the modified molecule, and the bioisostere pair representing the removal and insertion fragments.
 }
 \label{figure:data-preparation}
\end{figure*}

\subsection{Finding Molecular Pairs with Bioisosterism: Dataset construction}
\label{sec:dataset-generation}
In this section, we present our scheme for constructing a training dataset. It is a pivotal step in our work, where the dataset should be precisely tailored to align with the objective of DeepBioisostere.
To obtain a dataset with potential bioisosteric replacements, we begin from the ChEMBL database\cite{ChEMBL} and go through two steps of training dataset construction: pre-filtration and MMP identification.

\subsubsection{Pre-filtration}
\label{sec:pre-filtration}
We adopted a permissive pre-filtration strategy to maximally retain the data.
Simply, we excluded heavy molecules with molecular weights exceeding 800 $\mathrm{Da}$ and molecules consisting of salts.
Additionally, molecules with an inhibitory concentration greater than 10 $\mathrm{\mu{M}}$ are also excluded so that the remaining molecules can be regarded as possessing sufficient bioactivities.
This pre-filtration process provides a refined dataset appropriate to the next stage of MMP identification.
From the total of \fixme{4,182,178} ChEMBL molecules, we were able to obtain \fixme{1,054,335} molecules by pre-filtration.

\subsubsection{\fixme{Identifying Bioisosteric Replacements based on the BRICS MMPs}}
\label{sec:identifying-mmps-based-on-the-brics}
MMP is a pair of molecules in which all parts are identical except for a small fragment in each molecule.
From the pre-filtered ChEMBL database, we identified MMPs by fragmenting each molecule and then selecting molecule pairs that share common fragments.
It is crucial to adopt a proper fragmentation rule in this process since it determines the fragments corresponding to the \textit{difference} between each MMP, which are to be considered potential bioisosteres.
To consider more synthetically accessible modifications than the conventional strategy of simply cleaving rotatable bonds in the fragmentation process, we employed BRICS\cite{BRICS} rules for molecular fragmentation, implemented in the RDKit\cite{rdkit} package.

Using the BRICS rules, the ChEMBL molecules are represented as fragment-level graphs  (see \cref{figure:data-preparation}(a)).
Each node is \emph{irreducible fragment}, a subgraph of a molecule that cannot be further divided via fragmentation, where edges between nodes correspond to the BRICS bonds.
In this regard, a \emph{removal fragment} is defined as a single subgraph of a fragment-level molecular graph.
With these definitions, we enumerate multiple fragment combinations that involve a single removal fragment and remaining fragments for each BRICS-fragmented molecule.
The fragment combinations are further filtered with two criteria about their removal fragments: (1) a maximum of 12 heavy atoms in a removal fragment, and (2) the removal fragment should have fewer heavy atoms than the remaining fragments.

The aforementioned process results in an enormous number of fragment combinations derived from diverse molecules.
Then, the obtained fragment combinations are subsequently compared, where any molecular pairs with the same residual fragments are classified as MMPs along with their removal fragments.
We note that the majority of the resultant MMPs contain more than two matched fragment combinations.
For these cases, we exclusively selected the smallest removal fragments as bioisosteres (see \cref{figure:data-preparation}(b))
This selection of the smallest removal fragment distinguishes our training dataset from previous database-mining approaches that enumerate every bioisosteric pair, preventing the accumulation of redundant data in the training data. 
% \fixme{MMPs composed of molecules from different assays are filtered to leave only bioisosteric pairs.}
To ensure that only MMPs preserving bioactivity were considered, pairs with $|\Delta \mathrm{pChEMBL}| > 1$ were filtered out.
The final training dataset consists of a total of \fixme{2,874,580} MMPs specified about their corresponding bioisosteres.

Based on the final dataset, we collect the insertion fragments to build a \emph{fragment library}.
The fragment library is necessary for both the training and inference phases of DeepBioisostere as a source of insertion fragment candidates, which our model can utilize to optimize a given molecule.
For more details on the use of the fragment library of DeepBioisostere, refer to the \cref{sec:training_deepbioisostere,sec:molecular_optimization_process_with_deepbioisostere}.
Note that gathering all removal fragments will result in the same fragment library since each identified MMP results in two training data entries with reverse relationships in their removal and insertion fragments.
The resulting fragment library is composed of \fixme{140,096} diverse fragments.
Furthermore, we randomly split the fragment library at a ratio of 8:1:1, leading to \fixme{112,076}, \fixme{14,013}, and \fixme{14,007} fragments for training, validation, and test datasets, respectively.

\begin{figure*}[t!]
 \centering
 \includegraphics[width=0.9\textwidth]{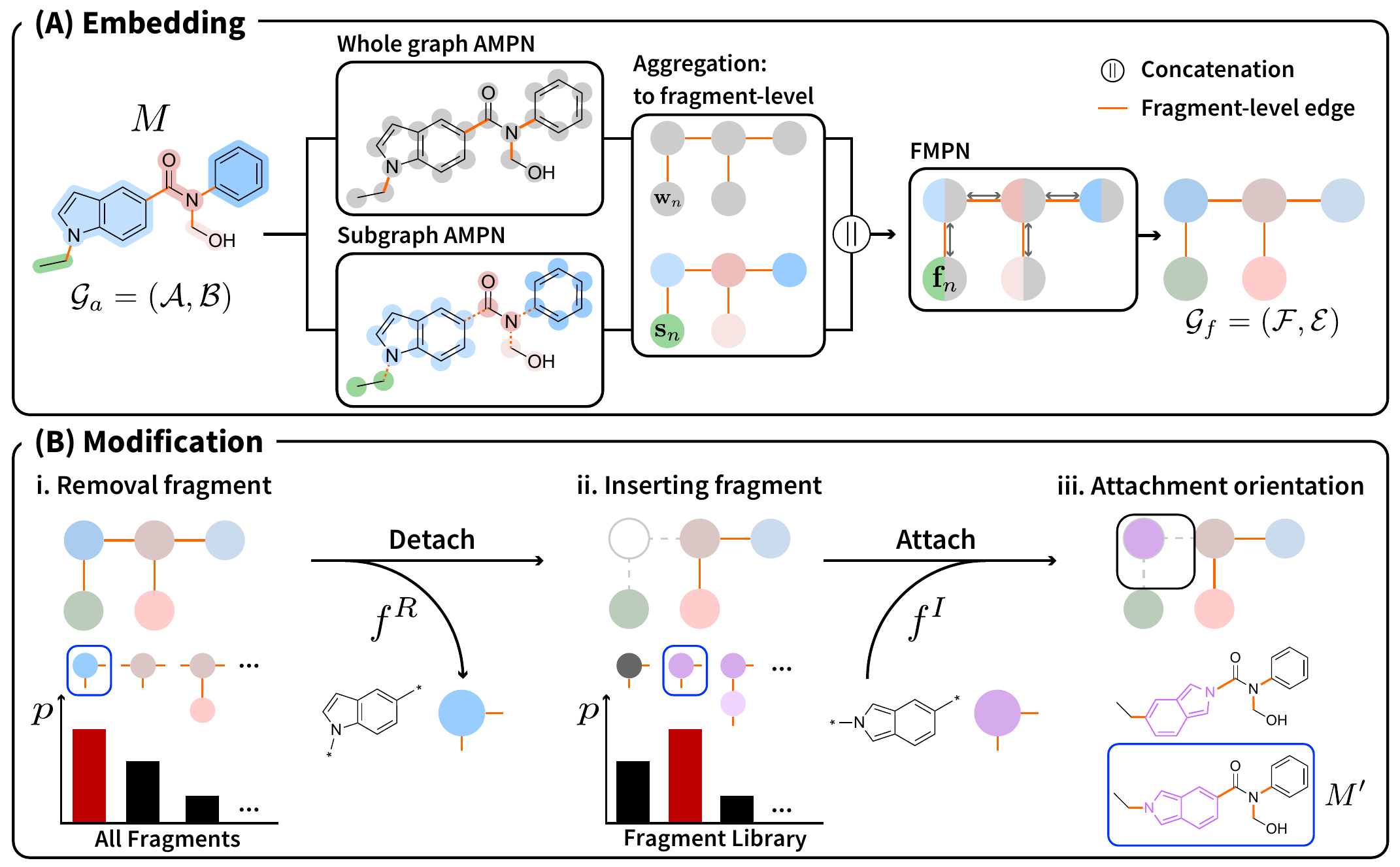}
 \caption{
 \textbf{Operation scheme of DeepBioisostere.}
 (a) Fragment-level feature embedding process.
 The original molecule, $M$, initially propagates through the whole graph and subgraph AMPN(atom message passing network).
 Subsequently, the atom-level embedding vectors are aggregated into a single fragment-level embedding vector, followed by an FMPN(fragment message passing network).
 (b) Based on the fragment-level vectors, DeepBioisostere sequentially selects the removal fragment, the insertion fragment, and the attachment orientation of the insertion fragment.
 The removal fragment is selected from all possible fragments of $M$, and the insertion fragment is chosen from our fragment library.
 }
\label{figure:model-architecture}
\end{figure*}

\subsection{DeepBioisostere: Deep Generative Modeling of Bioisosterism}
\label{sec:deepbioisostere-deep-generative-modeling}
We designed DeepBioisostere to optimize a molecule through three essential steps involving the selection of (1) the removal fragment, (2) the insertion fragment, and (3) the attachment orientation.
This sequential modeling was inspired by the traditional chemical modification process based on bioisosterism.
With this perspective, we define a chemical modification task as learning the joint distribution, $p\left(M,M'\textit{\textbf{C}}\right)$, of an original molecule, $M$, a modified molecular structure, $M'$, and a property control condition, $\textit{\textbf{C}}$.
We factorize the joint probability with $p\left(M,M',\textit{\textbf{C}}\right)=p\left(M,\textit{\textbf{C}}\right)\cdotp\left(M'\middle\vert{M},\textit{\textbf{C}}\right)$.
The joint probability $p\left(M,\textit{\textbf{C}}\right)$ is the likelihood of the combination of an original molecule to be modified and a property control objective for that molecule, arising from former experimental observations.
Therefore, we devise a generative model for the conditional probability, $p\left(M'\middle\vert{M},\textit{\textbf{C}}\right)$, implying the likelihood of a modified structure to achieve the given property control condition starting from an original structure.

Depending on the fragmentation rule, $\textbf{R}$, which defines what bond can be cleaved, the probability distribution must be different since achievable molecular structures from a single molecule would be different.
For example, if we do not allow the cleavage of the single bond connecting aromatic rings, we cannot obtain phenyl aniline by modifying a biphenyl compound.
Thus, we denote the objective conditional probability of our model as $p_{\textbf{R}}\left(M'\middle\vert{M},\textit{\textbf{C}}\right)$, reflecting the dependence of the chemical modification task on the bond cleavage rule, $\textbf{R}$.
For an MMP according to $\textbf{R}$, ($\textit{M}$,~$\textit{M}'$), their conditional probability can be reformulated by factorization with the conditional probabilities of corresponding three modification components as follows:

\begin{align*}
    p_{\textbf{R}}\left(M'\middle\vert{M},\textit{\textbf{C}}\right)
    &= p_{\textbf{R}}\left(f^R,f^I,A\middle\vert{M},\textit{\textbf{C}}\right) \\
    &= \underbrace{p_{\textbf{R}}\left(f^R\middle\vert{M},\textit{\textbf{C}}\right)}_{(1)}
    \cdot
    \underbrace{p_{\textbf{R}}\left(f^I\middle\vert{M},f^R,\textit{\textbf{C}}\right)}_{(2)} % \\
    \cdot
    \underbrace{p_{\textbf{R}}\left(A\middle\vert{M},f^R,f^I,\textit{\textbf{C}}\right)}_{(3)},
    \numberthis
    \label{eq:entire equation}
\end{align*}
where $f^R$ is a removal fragment, $f^I$ is an insertion fragment, and $A$ is an attachment orientation.
Note that for a modification of $M$ to $M'$, there might be more than one set of the three modification components, but we uniquely defined $f^R$ and $f^I$ to be only the smallest removal and insertion fragments in the training data construction procedure, respectively.
In this way, the ground truth for $p_{\textbf{R}}\left(f^R,f^I,A\middle\vert{M},\textit{\textbf{C}}\right)$ is defined to be zero if $f^R$ and $f^I$ are not the smallest ones to modify $M$ to another molecular structure, $M'$.
Since the smallest structural change can always be identified from any MMP, this choice does not impose any restriction on the probability distribution of $p_{\textbf{R}}\left(M'\middle\vert{M},\textit{\textbf{C}}\right)$.
The details about how we model each conditional probability of the \cref{eq:entire equation} are discussed in \cref{sec:modeling-sequential-modification}.

\subsubsection{Fragment-level Molecular Encoding}
\label{sec:molecule-encoding}

The overall framework of DeepBioisostere is depicted in \cref{figure:model-architecture}.
First, we encode an original molecule to be modified into fragment-level embedding vectors.
We adopt a hierarchical architecture to facilitate probability modeling at the fragment level.
An original molecule is represented in two ways: an atom-level graph $G_a$ and a fragment-level graph $G_f$.

\edit{We adopted the core architectural design of Modof proposed by \citet{Modof}, which demonstrates strong molecular property conditioning performance through a hierarchical framework.
In its design, as illustrated in \cref{figure:model-architecture}, atom-wise embeddings are first generated using an atom-level graph message passing network, then aggregated into fragment-level hidden vectors, which are subsequently processed by a fragment-level graph message passing network.
In DeepBioisostere, these two message passing networks are adapted for bioisosteric replacement and are referred to as the atom message passing network (AMPN) and fragment message passing network (FMPN), respectively.}

\edit{To tailor the model for bioisosteric replacement, we design the AMPN with two submodules having distinct parameters: a whole-graph AMPN operating on all molecular nodes and a subgraph AMPN operating only within individual fragments.
The subgraph AMPN is employed as a shared fragment embedder for both the original molecule and the fragment library, providing a consistent and comparable representation of molecular fragments.
This weight-sharing strategy ensures a comprehensive understanding of fragments and, in turn, allows the overall generative model to be formulated as a product of fragment-level conditional probabilities (\cref{eq:entire equation}).
In contrast to Modof, which decodes new fragments independently from separate latent vectors, DeepBioisostere evaluates candidate replacements by directly comparing the joint embedding of the removal fragment and each candidate insertion fragment from its library in this shared space.
Furthermore, when combined with our same-bioassay molecular pair dataset, this molecular encoding strategy enables the model to capture similarity in bioactivity as well as differences in other molecular properties.
We quantitatively analyze the contribution of this architectural design to multi-property optimization in Supplementary Information Section~5.1.
}

An initial atom feature vector, $\hat{\textbf{a}}$, is concatenated with the condition vector, $\textit{\textbf{C}}$, to yield an input atomic embedding vector by $\textbf{a}^{0}_{i} =[\hat{\textbf{a}}_{i}\Vert \textit{\textbf{C}}_i]$.
Details about constructing the initial atom feature vectors are summarized in the Supplementary Information.
We summarized key notations for model architecture in \cref{table:notation}.
Embedding vectors for both atoms and bonds are propagated through the AMPN layer as follows:
\begin{align}
    \textbf{m}^{t}_{i} &= \sum_{a_{j}\in \mathcal{N}(a_i)}{W_1^a\textbf{a}^{t}_j + W_2^a\textbf{b}_{ij}},\\
    \textbf{a}^{t+1}_{i} &=\textnormal{GRU}(\textbf{a}^{t}_{i}, \textbf{m}^{t}_{i}),
\end{align}
where $\mathcal{N}(a_i)$ is neighboring atoms of $a_i$, GRU is the gated recurrent unit introduced in \citet{GRU}
The whole graph AMPN operates on the entire molecular graph $\mathcal{G}_a$, updating atom-level embeddings based on the collective information of all atoms and bonds in the molecule.
In contrast, the subgraph AMPN works locally within each irreducible fragment, focusing its update on atoms only within the specific fragment, denoted as $\mathcal{A}(\hat{f})$.
This subgraph-based update strategy allows the model to learn the comprehensive features of each irreducible fragment for further hierarchical message passing.

\begin{table}[ht!]
    \centering
    
    \resizebox{0.8\textwidth}{!}{
    \renewcommand{\arraystretch}{1.1}
    \begin{tabular}{ll}
    
    \toprule
    \textbf{Notation} & \textbf{Description} \\
    \midrule
    $\mathcal{G}_a=(\mathcal{A},\mathcal{B})$ & Atom-level molecular graph \\
    $\mathcal{A}$, $\mathcal{B}$ & Set of atoms and bonds in $\mathcal{G}_a$, respectively \\
    $a\in\mathcal{A}$ & An atom in $\mathcal{G}_a$ \\
    $b_{ij}=\{a_i,a_j\}\in\mathcal{B}$ & A bond connecting atoms $a_i$ and $a_j$ in $\mathcal{G}_a$ \\
    $\mathbf{a}^t_i$ & An embedding vector of $a_i$ in $t^\mathrm{th}$ layer \\
    $\mathbf{b}_{ij}$ & An embedding vector of $b_{ij}$ \\

    \midrule

    $\mathcal{G}_f=(\mathcal{F},\mathcal{E})$ & Fragment-level molecular graph \\
    $\mathcal{F}$, $\mathcal{E}$ & \makecell[lt]{Set of irreducible fragments and edges in $\mathcal{G}_f$, \\ respectively} \\
    $f$ & A subgraph, fragmented by the BRICS rules \\
    $\hat{f}$ & An irreducible fragment \\
    $f^R$ & A \emph{removal} fragment \\
    $f^I$ & An \emph{insertion} fragment \\
    $e_{nm}\in\mathcal{E}$ & \makecell[lt]{An edge between irreducible fragments $\hat{f}_n$ and \\ $\hat{f}_m$ in $\mathcal{G}_f$, corresponding to a BRICS bond in $\mathcal{G}_a$} \\
    $\mathcal{A}(f)$ & A set of atoms included in a subgraph $f$ \\
    $\mathcal{A}(e_{nm})$ & A set of atoms composing $e_{nm}$ \\
    $\mathcal{F}(f)$ & A set of irreducible fragments included in a subgraph $f$ \\
    $\mathbf{f}^t_n$ & An embedding vector of $\hat{f}_n$ in $t^\mathrm{th}$ layer \\
    $\mathbf{e}_{nm}$ & An embedding vector of $e_{nm}$ \\
    $\mathbf{h}_f$ & An aggregated vector for $f$ \\
    $A$ & \makecell[lt]{An attachment orientation, a set of edges enumerated \\ based on the BRICS rules} \\

    \midrule
    
    $\left[\cdot\middle\Vert\cdot\right]$ & Element-wise concatenation \\
    $\sigma$ & Sigmoid activation function \\
    $W$ & Learnable weights \\
    
    \bottomrule
    
    \end{tabular}}
    \caption{
    Notations and the corresponding descriptions in this work
    }
    \label{table:notation}
\end{table}

The updated atomic embedding vectors are aggregated into fragment-level embedding vectors for each irreducible fragment as follows:
\begin{align}
    \textbf{w}_{n} &= \sum_{a_i\in \mathcal{A}(\hat{f}_n)}{\textbf{a}_i}, \label{eq:atom-aggregation}\\
    \textbf{e}_{nm} &= \sum_{a_i\in \mathcal{A}(e_{nm})}{\textbf{a}_i} \label{eq:bond-aggregation},
\end{align}
where $\mathbf{w}_n$ denotes a fragment-level embedding vector aggregated from the whole graph AMPN embeddings.
Atom-level embeddings from the subgraph AMPN are aggregated in the same way, which is denoted as $\mathbf{s}_n$.
Given that the subgraph AMPN updates the vectors without considering the information of adjacent atoms connected by the BRICS bonds, we exclusively employ the whole graph AMPN to aggregate the edge vectors of fragment-level graphs.
We concatenate the two aggregated fragment-level embedding vectors for further propagation in FMPN, which incorporates the feature vector of each fragment and intrinsically works like AMPN:
\begin{align}
    \textbf{m}_{\,n}^{\,t} &=\sum_{\hat{f}_m\in \mathcal{N}(\hat{f}_n)}{W_1^f\textbf{f}_{\,m}^{\,t} + W_2^f\textbf{e}_{\,nm}},\\
    \textbf{f}_{\,n}^{\,t+1} &=\mathrm{GRU}(\textbf{f}_{\,n}^{\,t}, \textbf{m}_{\,n}^{\,t}).
\end{align}

Here, $\textbf{f}_{\,n}^{\,0}=\left[ \textbf{w}_n\middle\Vert\textbf{s}_n \right]$, is an initial feature vector of an irreducible fragment.
The resulting fragment-level embedding vectors, $\left\{\textbf{f}_n\right\}$, are utilized for the modification of $M$.
Among the overall embedding process, the subgraph AMPN is significant since it allows DeepBioisostere to generate embedding vectors for previously unseen fragments.

With the fragment-level embedding vectors obtained by AMPN and FMPN, we model each conditional probability of \cref{eq:entire equation}.
During the substitution, both $f^R$ and $f^I$ can be a combination of multiple irreducible fragments.
Thus, we obtain the final embedding vector for each individual fragment through the aggregation of the involved irreducible fragments as follows:
\begin{equation}
    \mathbf{h}_f=\sum_{\hat{f}_n\in\mathcal{F}(f)} \mathbf{f}_n.
    \label{eq:sum-pooling}
\end{equation}
We note that $\mathbf{h}_f$ contains comprehensive information about the chemical structure of $f$, its relationships with surrounding fragments, and the given property control condition.

\subsubsection{Modeling Sequential Modification Steps}
\label{sec:modeling-sequential-modification}
With the aggregated embedding vectors, the DeepBioisotere model sequentially modifies the original molecule, $M$ (see \cref{figure:model-architecture}(B)).
First, the removal fragment selection module estimates the probability of removing each fragment.
This exclusively considers the allowed removal fragments, satisfying the criteria used during dataset construction, as detailed in \cref{sec:dataset-generation}.
For a potential removal fragment $f$, the likelihood is formulated as follows:
\begin{align}
    p\left(f\middle\vert{M},\textit{\textbf{C}}\right)
    \propto
    \sigma\big(\phi\left(\textbf{h}_f;\theta_R\right)\big), \label{eq:removal}
\end{align}
where $\phi(\cdot;\theta_R)$ is a removal fragment selection module parameterized with $\theta_R$.
After the removal fragment is determined, our insertion fragment selection module chooses an appropriate fragment from the predefined library established during dataset generation.
The following likelihood guides this selection:
\begin{align}
    p\left(f\middle\vert{M},f^R,\textit{\textbf{C}}\right) \propto \sigma\left(\phi\left(\left[\textbf{h}_{f^R}\middle\Vert\textbf{h}_f\right];{\theta_I}\right)\right). \label{eq:insertion}
\end{align}
Similarly, $\phi(\cdot;\theta_I)$ is an insertion fragment selection module parameterized with $\theta_I$.
Thus, the insertion fragment depends on the pre-selected removal fragment, its surroundings, and the given property control condition, which are comprehensively included in $\textbf{h}_{f^R}$.
\edit{This contrastive formulation enables the model to explicitly learn bioisosteric relationships between fragments.}

For an insertion fragment with two or more attachment sites, more than one \textit{attachment orientations} might be allowed.
For example, isoindole in the left side of \cref{figure:model-architecture}(B) can be attached to the remaining substructure in two different orientations.
In these cases, a proper attachment orientation should be determined to insert the selected fragment for desired property control.
We enumerate all attachment orientations that are allowed by the BRICS rules, and then the last module estimates their likelihood:
\begin{align}
    p\left(A\middle\vert{M},f^R,f^I,\textit{\textbf{C}}\right)
    \propto \prod_{e_{nm}\in A} \sigma\big(\phi\left(\textbf{e}_{nm};\theta_A\right)\big), \label{eq:attachment}
\end{align}
where $A$ is an attachment orientation, and $\phi\left(\cdot;\theta_A\right)$ is an attachment orientation selection module parameterized with $\theta_A$.
\cref{eq:attachment} indicates that each attachment orientation is evaluated based on the respective attaching sites (or edges) under the assumption that the compatibility of joining two fragments mainly depends on the local relationship.

\subsection{Training DeepBioisostere}
\label{sec:training_deepbioisostere}
As explained in section~\ref{sec:modeling-sequential-modification}, DeepBioisostere modifies a given molecule with three main modules.
To learn the conditional distribution $p\left(M'\middle\vert{M},\textit{\textbf{C}}\right)$ with them, we devised three loss functions for each module, $\mathcal{L}_{\textnormal{remove}}$, $\mathcal{L}_{\textnormal{insert}}$, and $\mathcal{L}_{\textnormal{attach}}$, respectively.
Then, we trained our model with their summation as the final training loss function.

The dataset constructed in \cref{sec:dataset-generation} includes only positive data of the ground-truth removal fragment, insertion fragment, and attachment orientation.
To make the removal fragment selection module likely select the positive removal fragment over others, we regard other allowed fragments of the original molecule as negative samples for each pair of bioisosteres in the dataset.
We employed the binary cross-entropy loss to formulate $\mathcal{L}_{\textnormal{remove}}$ as follows:
\begin{align}
    \mathcal{L}_{\textnormal{remove}} = -\log{p^R_{pos}} - \frac{1}{N_R} \sum_{f^R_{neg}}^{N_R}{\log{(1-p^R_{neg}})},  \label{eq:objective-removal}  
\end{align}
where $p^R_{pos}=\sigma(\phi(\textbf{h}_{f^R_{pos}};\theta_R))$, $p^R_{neg}=\sigma(\phi(\textbf{h}_{f^R_{neg}};\theta_R))$,  $f^R_{neg}$ is an allowed but negative fragment from the original molecule, and $N_R$ is the total number of allowed but negative fragments from the original molecule. 

The selection of insertion fragments is more complicated since our model should be able to retrieve the ground truth one among a vast number of potential ones from the fragment library. 
To avoid the inefficiency in evaluating tens of thousands of fragments for every data point, we employed the \textit{negative sampling} strategy inspired by \citet{SeoBBAR}
This strategy samples only a few fragments (20 fragments in this work) with the probability weakly proportional to their occurrence as negative samples.
With this negative sampling strategy, we formulated $\mathcal{L}_{\textnormal{insert}}$ as follows:
\begin{align}
    \mathcal{L}_{\textnormal{insert}} = -\log{p^I_{pos}} - \frac{1}{N_I} \sum_{f^I_{neg}}^{N_I}{\log{(1-p^I_{neg}})},  \label{eq:objective-inserting}  
\end{align}
where $p^I_{pos}=\sigma(\phi([\textbf{h}_R\Vert\textbf{h}_{f^I_{pos}}];\theta_I))$, $p^I_{neg}=\sigma(\phi([\textbf{h}_R\Vert\textbf{h}_{f^I_{neg}}];\theta_I))$, $N_I$ is the number of negative samples for each positive data.
By including negative samples in \cref{eq:objective-inserting}, we counteracted the bias towards always preferring more frequently occurring fragments over less frequent ones.
\edit{One potential concern is that, for a given molecular structure and its removal fragment pair, if there are two distinct ground-truth insertion fragments, one could mistakenly be sampled as a negative fragment during training.
However, since the final training objective is defined as the expectation over the entire dataset, the model will ultimately assign equal probabilities to both fragments.}

According to \cref{eq:attachment}, we estimate the probability of each attachment orientation as the product of the likelihood of the potential edges to be formed from the attachment.
Hence, we devised an edge-based objective function to train the attachment orientation selection module.
For each positive data, we enumerate all the potential edges allowed by the BRICS rules and classify them into positive and negative ones according to whether they are included in the ground-truth attachment orientation.
Then, we again set the binary cross-entropy loss in this task:
\begin{align*}
    \mathcal{L}_{\textnormal{attach}} &= -\frac{1}{N_{pos}}\sum_{e_{nm}^{pos}}^{N_{pos}}\log{p^A_{pos}}  \\
    &- \frac{1}{N_{neg}} \sum_{e_{nm}^{neg}}^{N_{neg}}{\log{(1-p^A_{neg}})}, \numberthis \label{eq:objective-attachment}  
\end{align*}
where $p^A_{pos}=\sigma(\phi(\textbf{e}_{nm}^{pos};\theta_A))$ and $p^A_{neg}=\sigma(\phi(\textbf{e}_{nm}^{neg};\theta_A))$.
$N_{pos}$ and $N_{neg}$ are the numbers of potential edges within the positive and negative orientations, respectively.

For training and validation, we split the whole dataset constructed in \cref{sec:dataset-generation} into training, validation, and test sets with a ratio of 8:1:1.
The dataset was split according to their insertion fragments so that the three sub-sets share no insertion fragment.
Thus, each dataset contains distinct bioisosteric replacements.
For the multi-property control, we used a total of four properties that can be readily computed based on the structure of a molecule: molecular weight, log$P$, QED,\cite{qed} and SAscore.\cite{sascore}
The training was finished in no more than 17 hours with a single NVIDIA RTX A4000 GPU for each property control condition.
Details about the training process can be found in the Supplementary Information.

\subsection{Molecular Optimization Process with DeepBioisostere}
\label{sec:molecular_optimization_process_with_deepbioisostere}
DeepBioisostere performs optimization with two inputs: a molecule to be improved and a property control condition.
The molecule is first analyzed with the BRICS rules to obtain its fragment-level graph, and all its substructures are enumerated.
Then, we filter the substructures based on the criteria used to identify MMPs in \cref{sec:identifying-mmps-based-on-the-brics}, allowing the modification to occur in a small part of the molecule.
Thus, a substructure that contains more than 12 heavy atoms or more heavy atoms than the other parts of the molecule is not considered in the removal fragment selection.
All the allowed fragments are examined by the removal fragment selection module, yielding likelihoods for each fragment.
Based on the multinomial distribution, removal fragments are sampled according to the desired number of modified structures.

For each selected removal fragment, DeepBioisostere explores the fragment library to find suitable insertion fragments for property control.
Since the chemical modification of a certain part should retain its surroundings, molecular fragments with different numbers of fragmented sites are not considered.
In addition, among the corresponding fragments, only those that satisfy the BRICS rules at the respective removal site of the original molecule are considered.
All the passed insertion fragments are then evaluated by the DeepBioisostere model, and the top insertion fragments are greedily selected. 
We note that, during the insertion fragment selection, either the occurrence data from the MMP database we created or from public chemical libraries, such as ChEMBL, is not utilized because the differences in occurrence are implicitly captured during the training process.

Lastly, the selected insertion fragments are examined in their attachment orientation.
The combinations between the fragmented sites of the insertion fragment and its surroundings are enumerated and then filtered using the BRICS rules.
The passed combinations are then prioritized by the DeepBioisostere model, and the insertion fragments are attached following the selected attachment orientations.
For more details about the generation process, refer to the Supplementary Information.

\backmatter

\section*{Data Availability}
The code for processing data from the ChEMBL database is available on GitHub: \url{https://github.com/Hwoo-Kim/DeepBioisostere.git}

\section*{Code Availability}
The code for training DeepBioisostere, sampling molecules, and evaluating the generated structures is available on GitHub: \url{https://github.com/Hwoo-Kim/DeepBioisostere.git}

\bibliography{sn-bibliography}% common bib file

@article{ScaffoldGVAE,
  title={ScaffoldGVAE: scaffold generation and hopping of drug molecules via a variational autoencoder based on multi-view graph neural networks},
  author={Hu, Chao and Li, Song and Yang, Chenxing and Chen, Jun and Xiong, Yi and Fan, Guisheng and Liu, Hao and Hong, Liang},
  journal={Journal of Cheminformatics},
  volume={15},
  number={1},
  pages={91},
  year={2023},
  doi={10.1186/s13321-023-00766-0},
  url = {https://doi.org/10.1186/s13321-023-00766-0}
}

@inproceedings{jin2020multi,
  title={Multi-objective molecule generation using interpretable substructures},
  author={Jin, Wengong and Barzilay, Regina and Jaakkola, Tommi},
  booktitle={International conference on machine learning},
  pages={4849--4859},
  year={2020},
  organization={PMLR}
}

@inproceedings{fu2021mimosa,
  title={Mimosa: Multi-constraint molecule sampling for molecule optimization},
  author={Fu, Tianfan and Xiao, Cao and Li, Xinhao and Glass, Lucas M and Sun, Jimeng},
  booktitle={Proceedings of the AAAI Conference on Artificial Intelligence},
  volume={35},
  number={1},
  pages={125--133},
  year={2021}
}

@article{loeffler2024reinvent,
  title={Reinvent 4: modern AI--driven generative molecule design},
  author={Loeffler, Hannes H and He, Jiazhen and Tibo, Alessandro and Janet, Jon Paul and Voronov, Alexey and Mervin, Lewis H and Engkvist, Ola},
  journal={Journal of Cheminformatics},
  volume={16},
  number={1},
  pages={20},
  year={2024},
  publisher={Springer}
}

@article{maziarka2020mol,
  title={Mol-CycleGAN: a generative model for molecular optimization},
  author={Maziarka, {\L}ukasz and Pocha, Agnieszka and Kaczmarczyk, Jan and Rataj, Krzysztof and Danel, Tomasz and Warcho{\l}, Micha{\l}},
  journal={Journal of Cheminformatics},
  volume={12},
  number={1},
  pages={2},
  year={2020},
  publisher={Springer}
}

@article{Bioisostere1,
	doi = {10.1002/minf.201000019},
	url = {https://doi.org/10.1002/minf.201000019},
	year = 2010,
	month = {may},
	publisher = {Wiley},
	volume = {29},
	number = {5},
	pages = {366--385},
	author = {Sarah R. Langdon and Peter Ertl and Nathan Brown},
	title = {Bioisosteric Replacement and Scaffold Hopping in Lead Generation and Optimization},
	journal = {Mol. Inf.}
}

@article{shan2020molopt,
  title={MolOpt: a web server for drug design using bioisosteric transformation},
  author={Shan, Jinwen and Ji, Changge},
  journal={Current computer-aided drug design},
  volume={16},
  number={4},
  pages={460--466},
  year={2020},
  publisher={Bentham Science Publishers direct}
}

@article{Bioisostere2,
	doi = {10.3390/ph13030036},
	url = {https://doi.org/10.3390/ph13030036},
	year = 2020,
	month = {feb},
	publisher = {{MDPI} {AG}},
	volume = {13},
	number = {3},
	pages = {36},
	author = {Alexej Dick and Simon Cocklin},
	title = {Bioisosteric Replacement as a Tool in Anti-{HIV} Drug Design},
	journal = {Pharmaceuticals}
}

@article{Bioisostere3,
	doi = {10.1021/acs.jafc.2c00785},
	url = {https://doi.org/10.1021/acs.jafc.2c00785},
	year = 2022,
	month = {may},
	publisher = {American Chemical Society ({ACS})},
	volume = {70},
	number = {36},
	pages = {11042--11055},
	author = {Xiaofeng Cao and Haiping Yang and Cheng Liu and Ruifeng Zhang and Peter Maienfisch and Xiaoyong Xu},
	title = {Bioisosterism and Scaffold Hopping in Modern Nematicide Research},
	journal = {J. Agric. Food Chem.}
}

@article{SimilarityPropertyPrinciple,
	doi = {10.1002/cjoc.201300390},
	url = {https://doi.org/10.1002/cjoc.201300390},
	year = 2013,
	month = {sep},
	publisher = {Wiley},
	volume = {31},
	number = {9},
	pages = {1123--1132},
	author = {Chaoqian Cai and Jiayu Gong and Xiaofeng Liu and Daqi Gao and Honglin Li},
	title = {Molecular Similarity: Methods and Performance},
	journal = {Chin. J. Chem.}
}

@misc{example_tetrazole1,
  title={Bioisosteres in drug discovery: focus on tetrazole},
  author={Zou, Yulin and Liu, Li and Liu, Junxiong and Liu, Guocheng},
  journal={Future Medicinal Chemistry},
  volume={12},
  number={2},
  pages={91--93},
  year={2020},
  publisher={Future Science}
}

@article{example_tetrazole2,
	doi = {10.1007/s10847-013-0334-x},
	url = {https://doi.org/10.1007/s10847-013-0334-x},
	year = 2013,
	month = {may},
	publisher = {Springer Science and Business Media {LLC}},
	volume = {78},
	number = {1-4},
	pages = {15--37},
	author = {Maqsood Ahmad Malik and Mohmmad Younus Wani and Shaeel Ahmed Al-Thabaiti and Rayees Ahmad Shiekh},
	title = {Tetrazoles as carboxylic acid isosteres: chemistry and biology},
	journal = {J Incl Phenom Macrocycl Chem}
}

@article{patani1996bioisosterism,
  title={Bioisosterism: a rational approach in drug design},
  author={Patani, George A and LaVoie, Edmond J},
  journal={Chemical reviews},
  volume={96},
  number={8},
  pages={3147--3176},
  year={1996},
  publisher={ACS Publications}
}

@article{ertl2007silico,
  title={In silico identification of bioisosteric functional groups.},
  author={Ertl, Peter},
  journal={Current opinion in drug discovery \& development},
  volume={10},
  number={3},
  pages={281--288},
  year={2007}
}

@article{kumari2020amide,
  title={Amide bond bioisosteres: Strategies, synthesis, and successes},
  author={Kumari, Shikha and Carmona, Angelica V and Tiwari, Amit K and Trippier, Paul C},
  journal={Journal of medicinal chemistry},
  volume={63},
  number={21},
  pages={12290--12358},
  year={2020},
  publisher={ACS Publications}
}

@article{seddon2018bioisosteric,
  title={Bioisosteric Replacements Extracted from High-Quality Structures in the Protein Databank},
  author={Seddon, Matthew P and Cosgrove, David A and Gillet, Valerie J},
  journal={ChemMedChem},
  volume={13},
  number={6},
  pages={607--613},
  year={2018},
  publisher={Wiley Online Library}
}

@article{InSilicoApproach,
  title = {In silico applications of bioisosterism in contemporary medicinal chemistry practice},
  volume = {3},
  ISSN = {1759-0884},
  url = {http://dx.doi.org/10.1002/wcms.1148},
  DOI = {10.1002/wcms.1148},
  number = {4},
  journal = {WIREs Computational Molecular Science},
  publisher = {Wiley},
  author = {Papadatos,  George and Brown,  Nathan},
  year = {2013},
  month = may,
  pages = {339–354}
}

@article{SwissBioisostere1,
  title={SwissBioisostere: a database of molecular replacements for ligand design},
  author={Wirth, Matthias and Zoete, Vincent and Michielin, Olivier and Sauer, Wolfgang HB},
  journal={Nucleic acids research},
  volume={41},
  number={D1},
  pages={D1137--D1143},
  year={2013},
  publisher={Oxford University Press}
}

@article{SimilarityApproach1,
  title={World Wide Web-based system for the calculation of substituent parameters and substituent similarity searches},
  author={Ertl, Peter},
  journal={Journal of Molecular Graphics and Modelling},
  volume={16},
  number={1},
  pages={11--13},
  year={1998},
  publisher={Elsevier}
}

@article{SimilarityApproach2,
	doi = {10.1021/ci990263g},
	url = {https://doi.org/10.1021/ci990263g},
	year = 1999,
	month = {dec},
	publisher = {American Chemical Society ({ACS})},
	volume = {40},
	number = {2},
	pages = {295--307},
	author = {Ansgar Schuffenhauer and Valerie J. Gillet and Peter Willett},
	title = {Similarity Searching in Files of Three-Dimensional Chemical Structures:{\hspace{0.167em}} Analysis of the {BIOSTER} Database Using Two-Dimensional Fingerprints and Molecular Field Descriptors},
	journal = {J. Chem. Inf. Comput. Sci.}
}

@article{SimilarityApproach3,
  title = {COSMOsim: Bioisosteric Similarity Based on COSMO-RS $\sigma$ Profiles},
  volume = {46},
  ISSN = {1549-960X},
  url = {http://dx.doi.org/10.1021/ci050464m},
  DOI = {10.1021/ci050464m},
  number = {3},
  journal = {Journal of Chemical Information and Modeling},
  publisher = {American Chemical Society (ACS)},
  author = {Thormann,  Michael and Klamt,  Andreas and Hornig,  Martin and Almstetter,  Michael},
  year = {2006},
  month = mar,
  pages = {1040–1053}
}

@article{SimilarityApproach4,
	doi = {10.1021/ci900085d},
	url = {https://doi.org/10.1021/ci900085d},
	year = 2009,
	month = {may},
	publisher = {American Chemical Society ({ACS})},
	volume = {49},
	number = {6},
	pages = {1497--1513},
	author = {Mike Devereux and Paul L. A. Popelier and Iain M. McLay},
	title = {Quantum Isostere Database: A Web-Based Tool Using Quantum Chemical Topology To Predict Bioisosteric Replacements for Drug Design},
	journal = {J. Chem. Inf. Model.}
}

@article{example_similarity,
	doi = {10.1021/ci0503964},
	url = {https://doi.org/10.1021/ci0503964},
	year = 2006,
	month = {jan},
	publisher = {American Chemical Society ({ACS})},
	volume = {46},
	number = {2},
	pages = {677--685},
	author = {Markus Wagener and Jos P. M. Lommerse},
	title = {The Quest for Bioisosteric Replacements},
	journal = {J. Chem. Inf. Model.}
}

@article{example_similarity2,
  title={Pharmacophore fingerprint-based approach to binding site subpocket similarity and its application to bioisostere replacement},
  author={Wood, David J and Vlieg, Jacob de and Wagener, Markus and Ritschel, Tina},
  journal={Journal of chemical information and modeling},
  volume={52},
  number={8},
  pages={2031--2043},
  year={2012},
  publisher={ACS Publications}
}

@article{example_structure,
  title={Bioisosteric similarity of molecules based on structural alignment and observed chemical replacements in drugs},
  author={Krier, Markus and Hutter, Michael C},
  journal={Journal of chemical information and modeling},
  volume={49},
  number={5},
  pages={1280--1297},
  year={2009},
  publisher={ACS Publications}
}

@misc{NumDescriptors,
  title={Handbook of molecular descriptors},
  author={Timmerman, Hendrik and Todeschini, Roberto and Consonni, Viviana and Mannhold, Raimund and Kubinyi, Hugo},
  year={2002},
  publisher={Weinheim: Wiley-VCH}
}

@article{DatabaseMiningApproach1,
  title={Extended Summary: BIOSTER—a database of structurally analogous compounds},
  author={Ujv{\'a}ry, Istv{\'a}n},
  journal={Pesticide Science},
  volume={51},
  number={1},
  pages={92--95},
  year={1997},
  publisher={John Wiley \& Sons, Ltd London}
}

@article{DatabaseMiningApproach2,
	doi = {10.1021/ci500282c},
	url = {https://doi.org/10.1021/ci500282c},
	year = 2014,
	month = {jul},
	publisher = {American Chemical Society ({ACS})},
	volume = {54},
	number = {7},
	pages = {1908--1918},
	author = {J{\'{e}}r{\'{e}}my Desaphy and Didier Rognan},
	title = {sc-{PDB}-Frag: A Database of Protein{\textendash}Ligand Interaction Patterns for Bioisosteric Replacements},
	journal = {J. Chem. Inf. Model.}
}

@article{DatabaseMiningApproach3,
  title={BoBER: web interface to the base of bioisosterically exchangeable replacements},
  author={Le{\v{s}}nik, Samo and {\v{S}}krlj, Bla{\v{z}} and Er{\v{z}}en, Nika and Bren, Urban and Gobec, Stanislav and Konc, Janez and Jane{\v{z}}i{\v{c}}, Du{\v{s}}anka},
  journal={Journal of Cheminformatics},
  volume={9},
  number={1},
  pages={1--8},
  year={2017},
  publisher={BioMed Central}
}

@article{MMP,
	doi = {10.1021/jm200452d},
	url = {https://doi.org/10.1021/jm200452d},
	year = 2011,
	month = {sep},
	publisher = {American Chemical Society ({ACS})},
	volume = {54},
	number = {22},
	pages = {7739--7750},
	author = {Ed Griffen and Andrew G. Leach and Graeme R. Robb and Daniel J. Warner},
	title = {Matched Molecular Pairs as a Medicinal Chemistry Tool},
	journal = {J. Med. Chem.}
}

@article{ChEMBL,
  title={ChEMBL: towards direct deposition of bioassay data},
  author={Mendez, David and Gaulton, Anna and Bento, A Patr{\'\i}cia and Chambers, Jon and De Veij, Marleen and F{\'e}lix, Eloy and Magari{\~n}os, Mar{\'\i}a Paula and Mosquera, Juan F and Mutowo, Prudence and Nowotka, Micha{\l} and others},
  journal={Nucleic acids research},
  volume={47},
  number={D1},
  pages={D930--D940},
  year={2019},
  publisher={Oxford University Press}
}

@incollection{drugproperty,
title = {Chapter 1 - Introduction},
editor = {Li Di and Edward H. Kerns},
booktitle = {Drug-Like Properties (Second Edition)},
publisher = {Academic Press},
edition = {Second Edition},
address = {Boston},
pages = {1-3},
year = {2016},
isbn = {978-0-12-801076-1},
doi = {https://doi.org/10.1016/B978-0-12-801076-1.00001-0},
url = {https://www.sciencedirect.com/science/article/pii/B9780128010761000010},
author = {Li Di and Edward H. Kerns}
}

@article{FindBioWithDNN,
  doi = {10.1021/acs.jcim.0c00290},
  url = {https://doi.org/10.1021/acs.jcim.0c00290},
  year = {2020},
  month = jun,
  publisher = {American Chemical Society ({ACS})},
  volume = {60},
  number = {7},
  pages = {3369--3375},
  author = {Peter Ertl},
  title = {Identification of Bioisosteric Substituents by a Deep Neural Network},
  journal = {Journal of Chemical Information and Modeling}
}

@article{DeepHop,
  doi = {10.1186/s13321-021-00565-5},
  url = {https://doi.org/10.1186/s13321-021-00565-5},
  year = {2021},
  month = nov,
  publisher = {Springer Science and Business Media {LLC}},
  volume = {13},
  number = {1},
  pages = {},
  author = {Shuangjia Zheng and Zengrong Lei and Haitao Ai and Hongming Chen and Daiguo Deng and Yuedong Yang},
  title = {Deep scaffold hopping with multimodal transformer neural networks},
  journal = {Journal of Cheminformatics}
}

@article{jin2019hierarchical,
  title={Hierarchical graph-to-graph translation for molecules},
  author={Jin, Wengong and Barzilay, Regina and Jaakkola, Tommi},
  journal={arXiv preprint arXiv:1907.11223},
  year={2019}
}

@inproceedings{jin2020hierarchical,
  title={Hierarchical generation of molecular graphs using structural motifs},
  author={Jin, Wengong and Barzilay, Regina and Jaakkola, Tommi},
  booktitle={International conference on machine learning},
  pages={4839--4848},
  year={2020},
  organization={PMLR}
}

@article{Modof,
  doi = {10.1038/s42256-021-00410-2},
  url = {https://doi.org/10.1038/s42256-021-00410-2},
  year = {2021},
  month = dec,
  publisher = {Springer Science and Business Media {LLC}},
  volume = {3},
  number = {12},
  pages = {1040--1049},
  author = {Ziqi Chen and Martin Renqiang Min and Srinivasan Parthasarathy and Xia Ning},
  title = {A deep generative model for molecule optimization via one fragment modification},
  journal = {Nature Machine Intelligence}
}

@article{TransformerMMP1,
  doi = {10.1186/s13321-021-00497-0},
  url = {https://doi.org/10.1186/s13321-021-00497-0},
  year = {2021},
  month = mar,
  publisher = {Springer Science and Business Media {LLC}},
  volume = {13},
  number = {1},
  pages = {},
  author = {Jiazhen He and Huifang You and Emil Sandstr\"{o}m and Eva Nittinger and Esben Jannik Bjerrum and Christian Tyrchan and Werngard Czechtizky and Ola Engkvist},
  title = {Molecular optimization by capturing chemist's intuition using deep neural networks},
  journal = {Journal of Cheminformatics}
}

@article{TransformerMMP2,
  doi = {10.1186/s13321-022-00599-3},
  url = {https://doi.org/10.1186/s13321-022-00599-3},
  year = {2022},
  month = mar,
  publisher = {Springer Science and Business Media {LLC}},
  volume = {14},
  number = {1},
  author = {Jiazhen He and Eva Nittinger and Christian Tyrchan and Werngard Czechtizky and Atanas Patronov and Esben Jannik Bjerrum and Ola Engkvist},
  pages = {},
  title = {Transformer-based molecular optimization beyond matched molecular pairs},
  journal = {Journal of Cheminformatics}
}

@article{TransformerADMET,
  doi = {10.1039/d2cp05332b},
  url = {https://doi.org/10.1039/d2cp05332b},
  year = {2023},
  publisher = {Royal Society of Chemistry ({RSC})},
  volume = {25},
  number = {3},
  pages = {2377--2385},
  author = {Lijuan Yang and Chao Jin and Guanghui Yang and Zhitong Bing and Liang Huang and Yuzhen Niu and Lei Yang},
  title = {Transformer-based deep learning method for optimizing {ADMET} properties of lead compounds},
  journal = {Physical Chemistry Chemical Physics}
}

@article{BRICS,
  doi = {10.1002/cmdc.200800178},
  url = {https://doi.org/10.1002/cmdc.200800178},
  year = {2008},
  month = oct,
  publisher = {Wiley},
  volume = {3},
  number = {10},
  pages = {1503--1507},
  author = {J\"{o}rg Degen and Christof Wegscheid-Gerlach and Andrea Zaliani and Matthias Rarey},
  title = {On the Art of Compiling and Using {\textquotesingle}Drug-Like{\textquotesingle} Chemical Fragment Spaces},
  journal = {{ChemMedChem}}
}

@article{SeoBBAR,
  title={Molecular Generative Model via Retrosynthetically Prepared Chemical Building Block Assembly},
  author={Seo, Seonghwan and Lim, Jaechang and Kim, Woo Youn},
  journal={Advanced Science},
  volume={10},
  number={8},
  pages={2206674},
  year={2023},
  publisher={Wiley Online Library}
}

@article{sascore,
  title={Estimation of synthetic accessibility score of drug-like molecules based on molecular complexity and fragment contributions},
  author={Ertl, Peter and Schuffenhauer, Ansgar},
  journal={Journal of cheminformatics},
  volume={1},
  pages={1--11},
  year={2009},
  publisher={Springer}
}

@misc{GRU,
  doi = {10.48550/ARXIV.1412.3555},
  url = {https://arxiv.org/abs/1412.3555},
  author = {Chung,  Junyoung and Gulcehre,  Caglar and Cho,  KyungHyun and Bengio,  Yoshua},
  title = {Empirical Evaluation of Gated Recurrent Neural Networks on Sequence Modeling},
  publisher = {arXiv},
  year = {2014},
  copyright = {arXiv.org perpetual,  non-exclusive license}
}

@article{phenyl-bioisosteres,
  doi = {10.1021/acs.jmedchem.1c01215},
  url = {https://doi.org/10.1021/acs.jmedchem.1c01215},
  year = {2021},
  month = sep,
  publisher = {American Chemical Society ({ACS})},
  volume = {64},
  number = {19},
  pages = {14046--14128},
  author = {Murugaiah A. M. Subbaiah and Nicholas A. Meanwell},
  title = {Bioisosteres of the Phenyl Ring: Recent Strategic Applications in Lead Optimization and Drug Design},
  journal = {Journal of Medicinal Chemistry}
}

@article{2019optimization,
  title={Optimization of molecules via deep reinforcement learning},
  author={Zhou, Zhenpeng and Kearnes, Steven and Li, Li and Zare, Richard N and Riley, Patrick},
  journal={Scientific reports},
  volume={9},
  number={1},
  pages={10752},
  year={2019},
  publisher={Nature Publishing Group UK London}
}

@article{qed,
  title={Quantifying the chemical beauty of drugs},
  author={Bickerton, G Richard and Paolini, Gaia V and Besnard, J{\'e}r{\'e}my and Muresan, Sorel and Hopkins, Andrew L},
  journal={Nature chemistry},
  volume={4},
  number={2},
  pages={90--98},
  year={2012},
  publisher={Nature Publishing Group UK London}
}

@misc{rdkit,
  title={RDKit: open-source cheminformatics. http://www.rdkit.org},
  author={Landrum, G}
}

@article{walters2018virtual,
  title={Virtual chemical libraries: miniperspective},
  author={Walters, W Patrick},
  journal={Journal of medicinal chemistry},
  volume={62},
  number={3},
  pages={1116--1124},
  year={2018},
  publisher={ACS Publications}
}

@article{gao2020synthesizability,
  title={The synthesizability of molecules proposed by generative models},
  author={Gao, Wenhao and Coley, Connor W},
  journal={Journal of chemical information and modeling},
  volume={60},
  number={12},
  pages={5714--5723},
  year={2020},
  publisher={ACS Publications}
}

@article{logp,
  title={Prediction of physicochemical parameters by atomic contributions},
  author={Wildman, Scott A and Crippen, Gordon M},
  journal={Journal of chemical information and computer sciences},
  volume={39},
  number={5},
  pages={868--873},
  year={1999},
  publisher={ACS Publications}
}

@inproceedings{pocket2mol,
  title={Pocket2mol: Efficient molecular sampling based on 3d protein pockets},
  author={Peng, Xingang and Luo, Shitong and Guan, Jiaqi and Xie, Qi and Peng, Jian and Ma, Jianzhu},
  booktitle={International Conference on Machine Learning},
  pages={17644--17655},
  year={2022},
  organization={PMLR}
}

@inproceedings{decompdiff,
  title={DECOMPDIFF: diffusion models with decomposed priors for structure-based drug design},
  author={Guan, Jiaqi and Zhou, Xiangxin and Yang, Yuwei and Bao, Yu and Peng, Jian and Ma, Jianzhu and Liu, Qiang and Wang, Liang and Gu, Quanquan},
  booktitle={Proceedings of the 40th International Conference on Machine Learning},
  pages={11827--11846},
  year={2023}
}

@article{deepicl,
  title={3D molecular generative framework for interaction-guided drug design},
  author={Zhung, Wonho and Kim, Hyeongwoo and Kim, Woo Youn},
  journal={Nature Communications},
  volume={15},
  number={1},
  pages={2688},
  year={2024},
  publisher={Nature Publishing Group UK London}
}

@inproceedings{targetdiff,
  title={3D Equivariant Diffusion for Target-Aware Molecule Generation and Affinity Prediction},
  author={Guan, Jiaqi and Qian, Wesley Wei and Peng, Xingang and Su, Yufeng and Peng, Jian and Ma, Jianzhu},
  booktitle={International Conference on Learning Representations},
  year={2023}
}

@article{lassalas2016structure,
  title={Structure property relationships of carboxylic acid isosteres},
  author={Lassalas, Pierrik and Gay, Bryant and Lasfargeas, Caroline and James, Michael J and Tran, Van and Vijayendran, Krishna G and Brunden, Kurt R and Kozlowski, Marisa C and Thomas, Craig J and Smith III, Amos B and others},
  journal={Journal of medicinal chemistry},
  volume={59},
  number={7},
  pages={3183--3203},
  year={2016},
  publisher={ACS Publications}
}

\section*{Acknowledgments}
% This work was supported by Basic Science Research Programs through the National Research Foundation of Korea(NRF), grant funded by the Ministry of Science and ICT(NRF-2023R1A2C2004376).
This work was supported by Basic Science Research Programs through the National Research Foundation of Korea, funded by the Ministry of Science and ICT (Grant Nos. RS-2023-NR077040,RS-2022-NR068758) and by the Ministry of Health and Welfare (Grant No. RS-2024-00512498) to H.K., S.M., W.Z., S.K., and W.Y.K.

\section*{Author contributions}
H.K. and S.M. equally contributed to this work.
H.K. and S.M. designed the methodology and carried out the experiments.
H.K., S.M., W.Z., and S.K. designed the experiments and analyzed the results.
All authors wrote the manuscript together, and W.Y.K. supervised the project.

\section*{Competing interests}
The authors declare no competing interests.

\includepdf[pages={1-},scale=1.0]{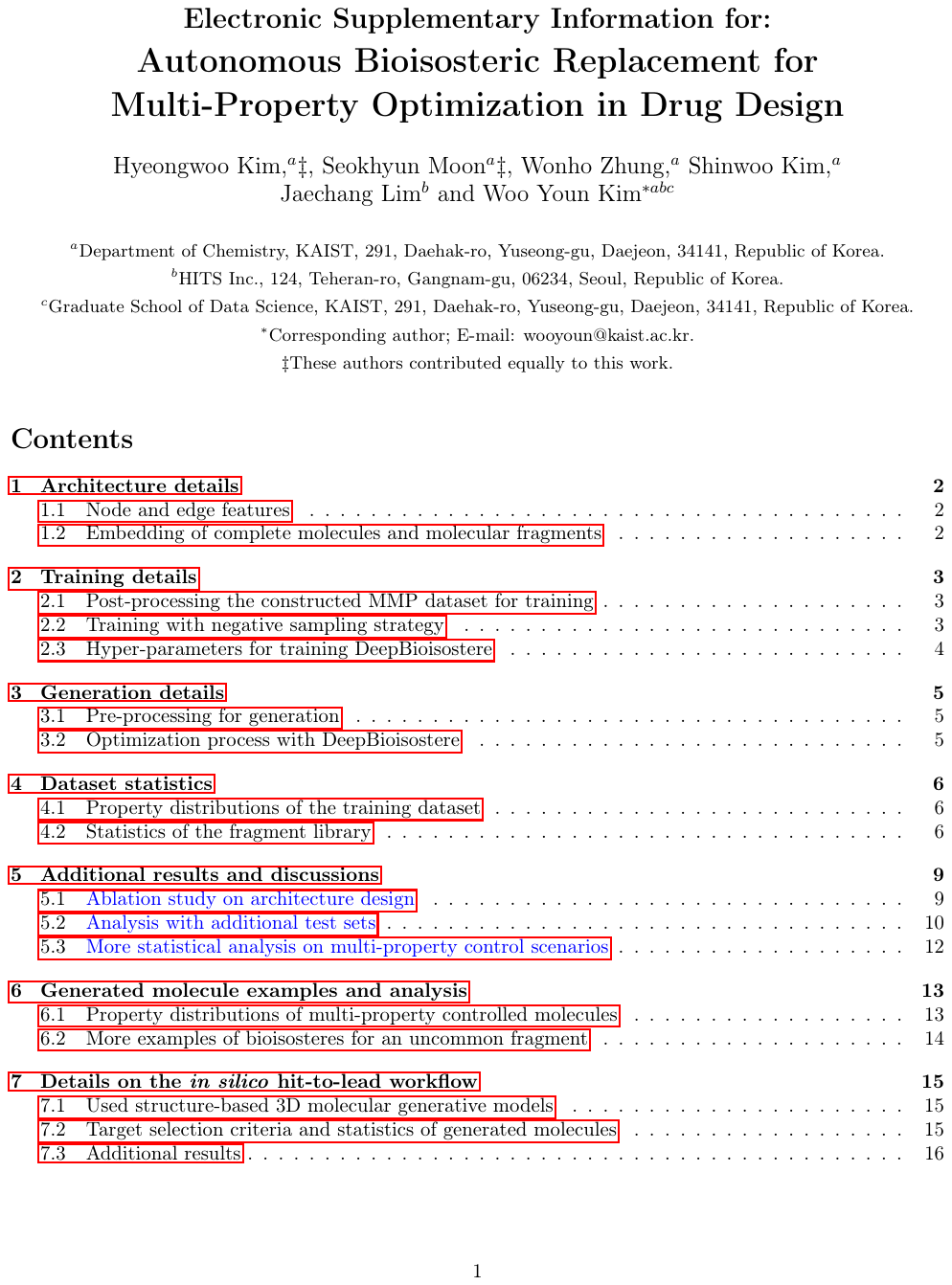}

\end{document}